\documentclass{aa}
	\usepackage{amssymb,amsfonts,amsmath,times,pict2e}
	\usepackage{natbib}
	\usepackage{graphicx}
	\usepackage{enumerate}
	\usepackage{subfigure}
	\usepackage{bm}
	\usepackage{amsmath,amsbsy}
	\usepackage{underscore}
	\usepackage{float} 
	\usepackage{mathtools}
	\usepackage{booktabs}
	\usepackage{multirow}
 	\usepackage{threeparttable}

	\usepackage{color}
	\usepackage{ulem}
	\definecolor{mygray}{gray}{0.6}
	\definecolor{magenta}{rgb}{0.858, 0.188, 0.478}
	
	\newcommand{\Rmnum}[1]{\expandafter\@slowromancap\romannumeral #1}

	\usepackage{xspace}
	\newcommand{\fg}[1]{Fig.~\ref{fig:#1}}
	\newcommand{\Fg}[1]{Figure~\ref{fig:#1}}

	\newcommand{\fgnum}[1]{\ref{fig:#1}}
	\newcommand{\eq}[1]{Eq.~(\ref{eq:#1})\xspace}
	\newcommand{\Eq}[1]{Equation~(\ref{eq:#1})\xspace}
	\newcommand{\eqs}[2]{Eqs.\ (\ref{eq:#1}) and (\ref{eq:#2})}
	\newcommand{\Eqs}[2]{Equations\  (\ref{eq:#1}) and (\ref{eq:#2})}

	\newcommand{\tb}[1]{Table~\ref{tab:#1}\xspace}
	\newcommand{\se}[1]{Sect.~\ref{sec:#1}\xspace}
	\newcommand{\Se}[1]{Section~\ref{sec:#1}\xspace}
	
	\newcommand{\ap}[1]{Appendix~\ref{ap:#1}\xspace}
	\newcommand{\aps}[2]{Appendix \ \ref{ap:#1} and \ref{ap:#2}}

	\newcommand{\ie}{i.e.}
	
	\newcommand{\eg}{e.g.}

         \newcommand{\Munit}{ \hat{M}}
	\newcommand{\AU}{ \  \rm AU}
	
	\newcommand{\Ms}{ \   M_\odot }
	\newcommand{\Msyr}{ \  M_\odot \,  \rm yr^{-1} }
	\newcommand{\yr}{ \   \rm yr}

	\newcommand{\Me}{ \  M_\oplus}

	\newcommand{\taus}{ \tau_{\rm s}}
	\newcommand{\tausf}{ \tau_{\rm s, F}}
	\newcommand{\tausb}{ \tau_{\rm s, B}}
	\newcommand{\tausd}{ \tau_{\rm s, D}}

	\newcommand{\Omegak}{\Omega_{\rm K}}

	\newcommand{\Sigmag}{\Sigma_{\rm g}}

	\newcommand{\alphag}{\alpha_{\rm g}}
	\newcommand{\alphat}{\alpha_{\rm t}}
	\newcommand{\Mp}{M_{\rm p}}

	\usepackage{CJKutf8}

\begin{document}

	\title{Pebble-driven Planet Formation around \\ Very Low-mass Stars and  Brown Dwarfs}

	\author{ Beibei Liu \inst{1} , Michiel Lambrechts \inst{1},  Anders Johansen\inst{1}, Ilaria Pascucci\inst{2} \and Thomas Henning \inst{3} }
	 \authorrunning{B. Liu et al.}
	\institute{
	Lund Observatory, Department of Astronomy and Theoretical Physics, Lund University, Box 43, 22100 Lund, Sweden \label{inst1} 
	\and
	Lunar and Planetary Laboratory, University of Arizona, Tucson, AZ 85721, USA \label{inst2} 
	\and Max-Planck-Institut f$\mathrm{ \ddot{u}}$r Astronomie, K$\mathrm{ \ddot{o}}$nigstuhl 17, 69117 Heidelberg, Germany \label{inst3} \\
	\email{bbliu@astro.lu.se}
	 }
	\date{\today}

	\abstract{
We conduct a pebble-driven planet population synthesis study to investigate the formation of planets around very low-mass stars and brown dwarfs, in the (sub)stellar mass range between $0.01\Ms$ and $0.1 \Ms$.  Based on the extrapolation of numerical simulations of planetesimal formation by the streaming instability, we obtain the characterisitic mass of the planetesimals and the  initial masses of the protoplanets (largest bodies from the planetesimal size distributions),
 in either the early self-gravitating phase or the later non-self-gravitating phase of the protoplanetary disk evolution. 
We find that the initial protoplanets form with masses that increase with host mass, orbital distance and decrease with age.  
Around  late M-dwarfs of $0.1\Ms$, these protoplanets can grow up to Earth-mass planets by pebble accretion. However, 
around brown dwarfs of $0.01\Ms$, planets do not  grow larger than Mars mass when the initial protoplanets are born early  in self-gravitating disks, and their growth stalls at around $0.01$ Earth-mass when they are born late in non-self-gravitating disks. Around these low mass stars and brown dwarfs we find no channel for gas giant planet formation, because the solid cores remain too small.
When the initial protoplanets form only at the water-ice line, the final planets typically have ${\gtrsim} 15\%$ water mass fraction.
Alternatively, when the initial protoplanets form log-uniformly distributed over the entire protoplanetary disk,
 the final planets are either very water-rich (water mass fraction ${\gtrsim}15\%$) or entirely rocky (water mass fraction ${\lesssim}5\%$). } 


	\maketitle

\section{Introduction}
\label{sec:introduce}

 Brown dwarfs are substellar objects in the mass range between $13$ and $80$ Jupiter masses (approximately $0.013$ and $0.08 \Ms$). They fall below the stable hydrogen-burning mass limit but can sustain deuterium and lithium nuclear fusion. These brown dwarfs, together with very low-mass hydrogen-burning M dwarf stars of masses ${\lesssim}0.1\Ms$ (effective temperature below $2700$ K, stellar type later than M$7$) are referred to as ultra-cool dwarfs (UCDs). 
  Noticeably, the UCDs represent a significant fraction (${\sim} 15\%{-}30\%$) of stars in the galaxy  \citep{Kroupa2001,Chabrier2003,Kirkpatrick2012,Muzic2017}. 
   
The evidence for dust disks around young brown dwarfs has first been probed by the presence of an infrared excess in the spectral energy distribution (SED) \citep{Comeron2000,Natta2001,Pascucci2003}, and later by the detection of (sub)millimeter emission \citep{Klein2003} from the cold dust.
Later, both dust continuum emission and $\rm CO$ molecular line emission have been observed for a large sample of brown dwarf disks through surveys with the Atacama Large Millimeter/submillimeter Array (ALMA) \citep{Ricci2012,Ricci2014,Testi2016}
Based on the infrared spectroscopic surveys by the Spitzer space telescope, \cite{Apai2005} found that grain growth, crystallization, and vertical settling all take place in such young brown dwarf disks.  
Furthermore, multi-wavelength observations resolve low spectral indexes for these brown dwarf disks, implying dust grains already grow up to mm/cm in size \citep{Ricci2010,Ricci2014,Ricci2017}  \footnote{ The above spectral index interpretation relies on the assumptions that the dust emission is optically thin and the opacity is dominated by absorption rather than scattering at the observed wavelengths   \citep{DAlessio2001,Draine2006,Zhu2019}. Recently, \cite{CarrascoGonzalez2019} fitted the SED of HL Tau disk from  ALMA and Very Large Array (VLA) data, by accounting for both scattering and absorption in dust opacity and neglecting any underlying assumption on the optical depth.  From this analysis, they found that the grains have indeed already grown to  millimeter size.}.
All these mentioned brown dwarf disk studies, combined with the vast literature on disks around more massive stars \citep{Przygodda2003,Natta2004,Perez2015,Tazzari2016}, convincingly support that the robustness of the first step of planet formation, grain growth, is ubiquitous among different young (sub)stellar environments, extending down to the brown dwarf regime.

Spectroscopic measurements of mass accretion rates onto brown dwarfs have found $\dot M_{\rm g}$ of order of  $10^{-12} \Msyr$ \citep{Muzerolle2005,Mohanty2005,Herczeg2009}, much lower than the nominal value for solar-mass stars ($10^{-8} \Msyr$) \citep{Hartmann1998,GarciaLopez2006,Manara2012}, implying a decreasing trend of mass accretion rate with (sub)stellar mass down to the brown dwarf regime.  Brown dwarfs also have protoplanetary disks of relatively low mass \citep{Andrews2013,Mohanty2013}.  Recent studies plausibly suggest a lower disk-to-star mass ratio ($M_{\rm d}/M_{\star}$) for disks around brown dwarfs and very low-mass stars compared to those around T Tauri stars \citep{Harvey2012,Pascucci2016,Testi2016}. 

 The lifetime of disks around M dwarfs and brown dwarfs are typically longer compared to  high-mass stars \citep{Carpenter2006,Scholz2007,Riaz2012}.  The stellar multiplicity however exhibits the opposite trend. The binary frequency is ${\sim}10\%$ for brown dwarfs, and this fraction increases with the stellar mass, up to ${\sim} 50\%$ for solar-type stars \citep{Burgasser2006,Ahmic2007,Lafreniere2008,Joergens2008,Raghavan2010,Fontanive2018}. 
 
Young solar-mass stars contract over a few tens of Myr to settle down to the main-sequence where they initiate nuclear fusion. However, the Kelvin-Helmholtz contraction timescale for UCDs is much longer ($100$ Myr to a few Gyr, \cite{Chabrier1997}). As UCDs contract and cool down, their luminosities become more than two orders of magnitude fainter \citep{Chabrier1997,Burrows1997}. 
Eventually, they stop shrinking when the gas is dense enough that electrons become degenerate in the interior.  Besides, these UCDs are also faster rotators, with a typically rotation period of $0.5{-}1$ day \citep{Herbst2007}. The magnetic B-field strengths of these low-mass objects are comparable to those of T Tauri stars, reaching kG level \citep{Reiners2012}, while their magnetic activities can last longer compared to those of T Tauri stars \citep{Reiners2010}.

Current exoplanet surveys, such as the MEarth project \citep{Nutzman2008}, TESS (Transiting Exoplanets Survey Satellite, \cite{Ricker2015}), SPECULOOS (search for Habitable planets Eclipsing Ultra-cool Stars, \cite{Gillon2013,Burdanov2018}),  CARMENES (Calar Alto high-Resolution search for M dwarfs with Exoearths with Near-infrared and optical Echelle Spectrographs, \cite{Quirrenbach2018}) and Project EDEN (ExoEarth Discovery and Exploration Network) transit survey, are currently operational to probe which types of planets and how frequent they are around stars in the ultra-cool dwarf regime. The planets discovered around TRAPPIST-1 \citep{Gillon2016,Gillon2017} and Teegarden’s star \citep{Zechmeister2019} are two known  systems in these pioneering  explorations.  These two hosts are both late M-dwarfs, and there are no planets detected by radial velocity/transit around brown dwarfs yet.

 From the theoretical point of view, the fundamental question to address is whether and how planets form in these low-mass (sub)stellar systems.  As stated, various host and disk properties for UCDs exhibit certain similarities and diversities with high mass stars.  
 In \cite{Liu2019b} we constructed  a pebble-driven planet formation model and focused  on the growth and migration of planets around stars from late M-dwarfs to solar mass.  We found a linear correlation between the characteristic planet mass (the maximum planet core mass) and the mass of the stellar host. 
On the other hand, pebble accretion is also proposed to explain the formation of giant planet satellites \citep{Shibaike2019,Ronnet2020}. 
 In this work we extend our population synthesis study to a lower (sub)stellar mass range, from $0.01 \Ms$ (${\approx} 10 \  M_{\rm Jup}$) to $0.1\Ms$. We present the first results of population synthesis simulations for planet formation around ultra-cool dwarfs from the pebble accretion perspective. Specifically, we aim to explore  how the planet mass correlates with the host's mass in this regime.    
 
The paper is designed as follows. \Se{model} briefly reviews our previous model \citep{Liu2019b} and describes three major improvements from this work.  The growth tracks of individual protoplanets  (planet mass vs semi-major axis) around UCD disks are illustrated in \se{growth}.  We simulate the growth and migration of a large number of protoplanets by Monto Carlo sampling their initial conditions.  In \se{pps}  we present the resulting planets with their masses, semi-major axes and water fractions.  We summarize our findings and discuss the implications in \se{conclusion}.

	\section{Planet formation model}
	\label{sec:model}

	\cite{Liu2019b} constructed a pebble-driven planet formation population synthesis model. Details of the model can be found in their Section 2. The most important equations of the disk model are recapitulated in \ap{diskmodel}. 	
	In this study we include three major improvements compared to the previous one. In \se{disk} we introduce the new adopted disk accretion rate  based on observations and their implication.  The dependences of the luminosity on time  and (sub)stellar mass are considered in \se{stellar}. In \se{embryo} the  initial mass of the protoplanets is now based on the extrapolation of  planetesimal masses from streaming instability simulations presented in literature. The pebble size is adopted to be the same as \cite{Liu2019b}, which is evaluated in \se{dust}.  
In particular, we emphasize the influence of the host's mass on these key physical quantities.

	\subsection{Disk properties}
	\label{sec:disk}
	
		\begin{figure}[t]
	     \includegraphics[width=\columnwidth]{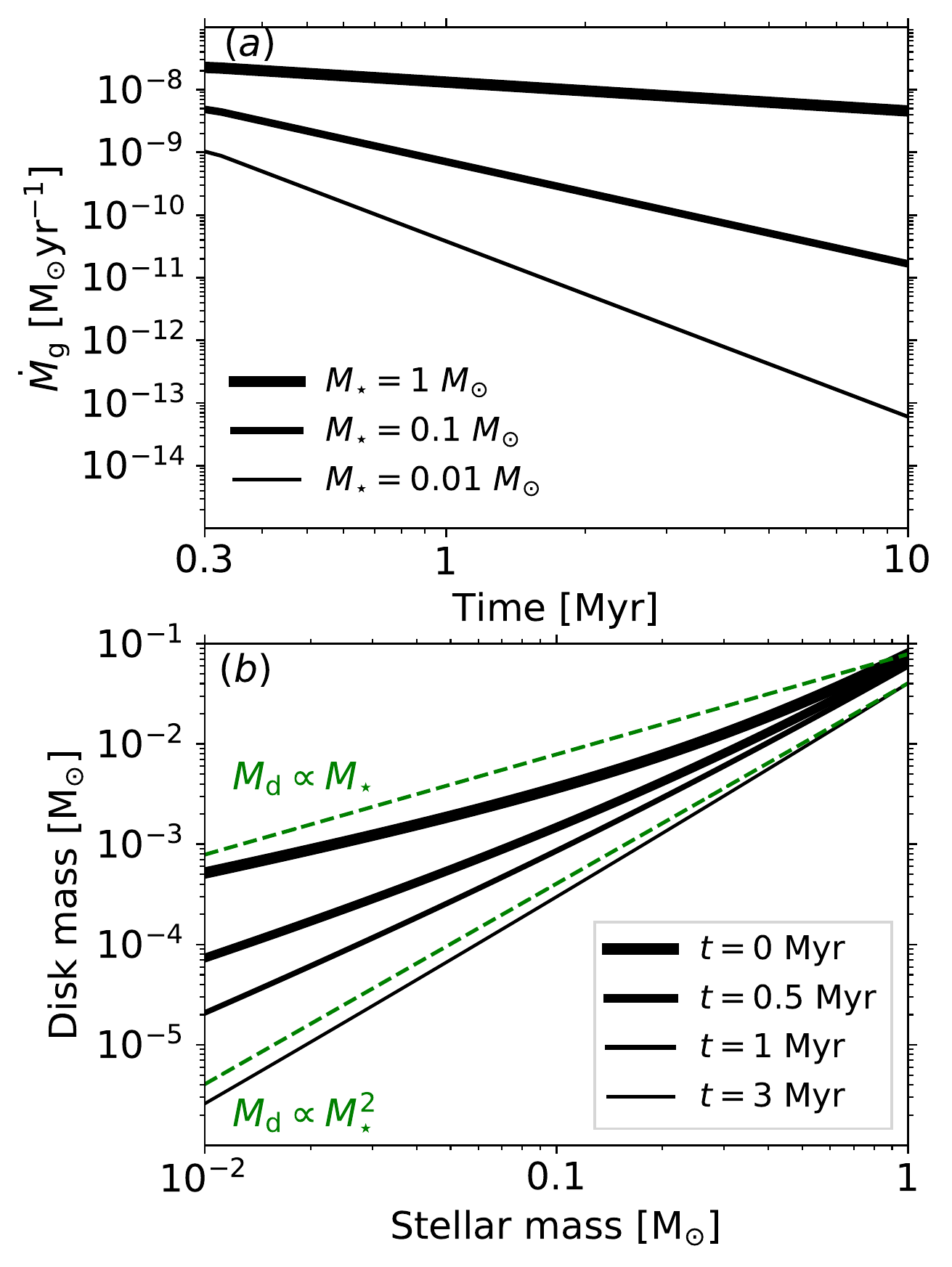}
		\caption{ Panel (a): time evolution of disk accretion rates among three different mass stars, based on Eq. (6) of \cite{Manara2012}. The disk accretion rate drops faster with decreasing mass of the central object. 
		Panel (b): Disk mass vs (sub)stellar mass at different times, based on $ \dot M_{\rm g}$ measurements of \cite{Manara2012}. The solid lines from thick to thin represent the ages of systems, from $0$, $0.5$, $1$ to $3$ Myr.  The green dashed lines correspond to the linear and  quadratic relation between the disk mass and the stellar mass.  When the disk  is ${\lesssim}1$ Myr old, $M_{\rm d} {\propto} M_{\star}$, while $M_{\rm d} {\propto} M_{\star}^2$ when the age is larger than ${\sim}1{-}3$ Myr. 
        }
	\label{fig:diskm}
	\end{figure}

	 An analytical self-similar solution for the viscous disk evolution  \citep{Lynden-Bell1974,Hartmann1998} was used in our previous work. Alternatively, here we adopt a fitting formula of $\dot M_{\rm g} (M_{\star}, t)$ from \cite{Manara2012}, based on observations of a large sample of sources in Orion Nebula Cluster, which gives  
	\begin{equation}
	\begin{split}
	  \log\left[\frac{\dot M_{\rm g}}{\Msyr} \right]  = &  -5.12 -0.46  \log\left[\frac{t}{\yr}\right] 
	  -5.75\log\left[\frac{ M_{\star}}{\Ms}\right]  \\
	 &  +1.17 \log\left[\frac{t}{\yr}\right]\log\left[\frac{ M_{\star}}{\Ms}\right],  
	 \end{split}
	 \label{eq:mdot}
	\end{equation}
	where $M_{\star}$ is the mass of central host, and $t$ is the age of the disk.
	Since the age determination for young protostars is highly uncertain during the early infall stage, as pointed out by \cite{Manara2012}, \eq{mdot} is invalid when the disk is younger than $0.3$ Myr.   We  assume that $\dot M_{\rm g}$ remains a constant when $t{<}0.3 \rm \ \rm Myr$.  In their sample, the mass of the central host ranges from $2\Ms$ to $0.05\Ms$. We assume here that the above equation can be extrapolated down to the very low-mass brown dwarf of $0.01\Ms$.  
	
	In contrast to this approach,   \cite{Liu2019b} assumed a viscously evolving disk, where $\dot M_{\rm g}$ depends on time as  $t^{{\approx}-3/2}$. Although several studies  have attempted to link the observed disks with viscous accretion theory,  the connection is not firmly established yet (\eg , \citealt{Lodato2017,Mulders2017,Najita2018}). 
	  Nonetheless, the disk evolution $\dot M_{\rm g} (M_{\star}, t)$ from this work is purely based on observations. 

	\fg{diskm}a shows the evolution of disk accretion rates for central hosts of $0.01\Ms$ (${=}10\ M_{\rm Jup}$), $0.1 \Ms$ and $1 \Ms$, respectively. There are two distinctive features. First, disks around less massive stars have lower accretion rates.  Second and more importantly, as stated by \cite{Manara2012}, disk accretion rates around less massive stars also decrease more rapidly. We also note that,  based on disk accretion rates measured from  X-shooter spectrograph, more recent studies suggested a steeper $\dot M_{\rm g}{-}M_{\star}$ for lower mass stars \citep{Manara2017,Alcala2017}, which seems in agreement with faster disk evolution around these stars.  We find in \fg{diskm}a that  at the beginning $\dot M_{\rm g} {=}10^{-9} \Msyr$ in a brown dwarf disk, which is  a factor of $5$ and $30$ times lower than the accretion rates in  $0.1 \Ms$ and $1 \Ms$ star disks. However, after $2$ Myr the disk accretion rate in the brown dwarf disk drops by more than two orders of magnitude, while $\dot M_{\rm g} \simeq 10^{-8} \Msyr$ for the solar-mass star, which is only reduced by a factor of $3$ compared to its initial value.

We can calculate the gas disk mass ($M_{\rm d}$) by integrating $\dot M_{\rm g}$ over time (choose $t{=}10$ Myr) for the host of specific $M_{\star}$.
	By doing so, we need to assume that the disk material is mainly accreted onto the central star, rather than carried away by a disk wind. It is true for viscous disks where most gas in the disk is accreted inward while a small amount of gas spreads far away to conserve the angular momentum \citep{Lynden-Bell1974}. Therefore,  we can link the mass accretion rate for the inner disk region to the whole disk mass. 
	
	 The results are illustrated in \fg{diskm}b. We find that at the early stage the disk mass tends to scale linearly with the mass of the central host. However, since the disk around low-mass stars/brown dwarfs evolves rapidly,  at $t {\sim} 1{-}3$ Myr the disk mass is roughly proportional to the host's mass square ($M_{\rm d}{ \propto} M_{\star}^{2}$). This inferred stellar mass square dependence  also matches the measurements of disk accretion rates in the Chamaeleon I star forming region \citep{Manara2016}, which have similar ages of $1{-}3$ Myr. 
	  The early linear $M_{\rm d}{-}M_{\star}$ correlation ($t{\lesssim} 0.5$ Myr) could be partially due to the constant $\dot M_{\rm g}$ assumption in the short embedded phase. Nevertheless, the fact that $M_{\rm d}{-}M_{\star}$ relation increases steeply with time always holds as long as the disk evolves more rapidly for very low-mass stars/brown dwarfs than that for higher-mass stars.
	We also note that such time-dependence is not only observed for the gas component (inferred from $\dot M_{\rm g}$). There is clear evidence that the correlation between the dust mass and stellar mass becomes steeper with time as well \citep{Pascucci2016,Ansdell2017}. 
	 

	\begin{figure}[t]
	     \includegraphics[width=\columnwidth]{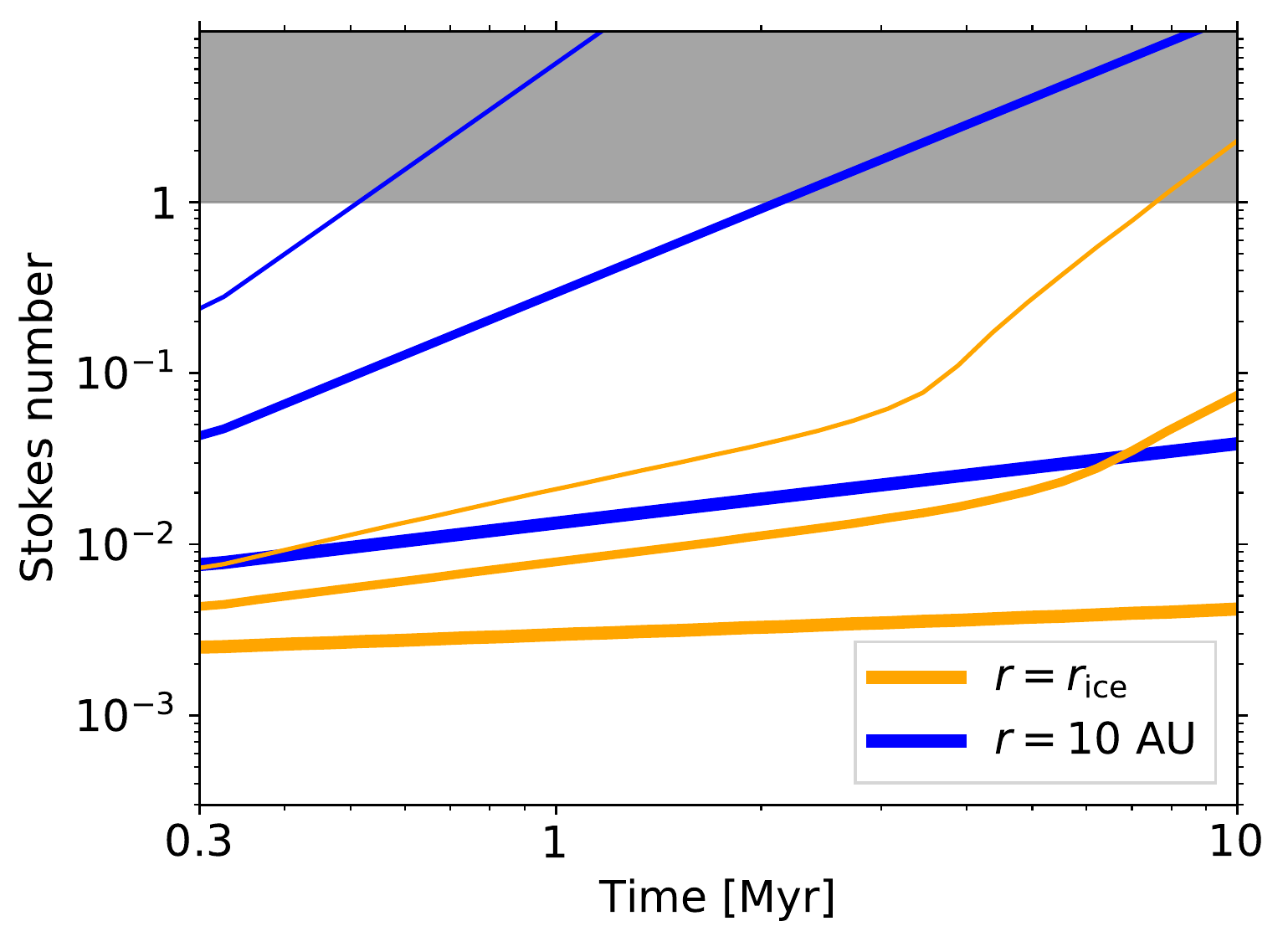}
		\caption{ Time evolution of Stokes numbers of mm-sized particles at $r= r_{\rm ice}$ (orange) and at $r= 10$ AU (blue) among three different mass hosts.   The thickness of lines represents the masses of the hosts, $1\Ms$, $0.1\Ms$ and $0.01\Ms$, respectively.	
 The Stokes number quickly gets larger than unity at the distant disk location around low-mass stars, resulting in inefficient pebble accretion.  
        }
	\label{fig:St}
	\end{figure}
		\begin{figure}[t]
	     \includegraphics[width=\columnwidth]{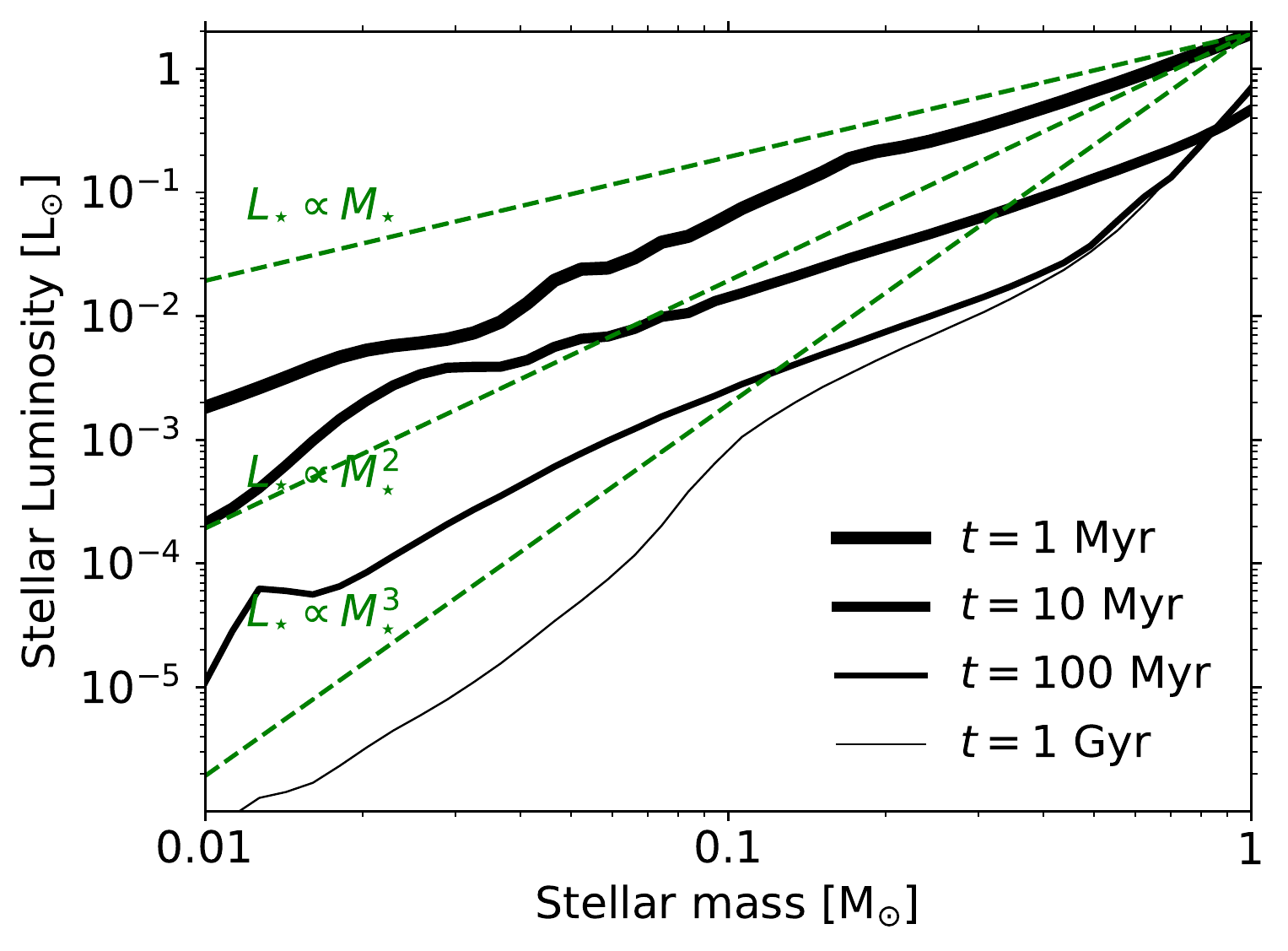}
		\caption{  Time and (sub)stellar mass dependencies on the luminosity of the host  based on the evolutionary model of \cite{Baraffe2003,Baraffe2015}. The solid lines from thick to thin indicate the systems' ages from  $1$ Myr,  $10$ Myr, $100$ Myr to $1$ Gyr.   The green dashed lines correspond to a linear, quadratic, cubic and fourth power correlation between the luminosity and mass. We find that when the central host is younger than $10$ Myr, $L_{\star}{\propto} M_{\star}^{1-2}$, while   $L_{\star} {\propto} M_{\star}^{3-4}$ when hosts are generally older than $10$ Myr.
		}
	\label{fig:Lstar}
	\end{figure}
	
 \subsection{Dust growth}
 \label{sec:dust}

    Similar to \cite{Liu2019b}, we also assume in this paper that dust grains can efficiently grow to millimeter size.  This characteristic size is motivated from two aspects. Firstly, the millimeter spectral index measured for the disks around various mass stars (including brown dwarfs) are much lower than the typical interstellar medium (ISM) value. It indicates that the grains already grow to such sizes in protoplanetary disks over a wide range of ages \citep{Draine2006,Ricci2014,Perez2015,Tazzari2016,Ricci2017,Pinilla2017}. Secondly, \cite{Zsom2010} studied the coagulation of silicate dust aggregates in protoplanetary disks based on the collision outcomes from laboratory experiments \citep{Guttler2010}. They found that particles stall at roughly millimeter size due to the bouncing barrier. The sticking of particles is determined by the surface energy, which correlates with the material properties. Recently, \cite{Musiolik2019} reported that the surface energy of ice aggregates is comparable to that of silicates when the disk temperature is lower than $180$ K. Therefore, the growth pattern and the bouncing-limited size would be quite similar for these two types of dust grains.
    
Rather than build a sophisticated model for dust growth and radial drift \citep{Birnstiel2012}, we here assume that the pebble mass flux ($\dot M_{\rm peb}$) generated from dust reservoir in the outer disk regions is attached to the gas flow with a constant mass flux ratio,  such that  $\xi {\equiv} \dot M_{\rm peb}/ \dot M_{\rm g} $ does not change with time \citep{Johansen2018}. Particles of very low Stokes number are well-coupled to the disk gas. The gas and dust particles drift inward together with the same velocities, and therefore $\xi$ remains equal to the initial disk metallicity.  When the Stokes number is high, pebbles radially drift faster than the disk gas. In this case, in order to maintain a constant flux ratio, $\Sigma_{\rm peb}/\Sigma_{\rm g}$ becomes lower than the nominal disk metallicity (the pebble surface density is reduced). The millimeter-sized pebbles in disks around GK and early M-dwarfs are typically in the former circumstance, while such pebbles in disks around very low-mass stars and brown dwarfs could be in the latter circumstance, especially at the late evolutional stage.

 We would like to point out that the above constant mass flux ratio assumption is a global concept. Even thoygh the global $\Sigma_{\rm peb}/\Sigma_{\rm g}$ is no higher than the disk metallicity under this assumption,  the density of the solids can still be enriched at local disk locations due to a variety of mechanisms. For instance, several studies proposed that the solid-to-gas ratio can be enhanced at the water-ice line \citep{Ros2013,Schoonenberg2017,Drazkowska2017}. This is because the water-vapor is released by the inward drifting icy pebbles when these pebbles cross the ice line.   The water-vapor diffuses back to the low density region exterior of the ice line  and re-condenses  onto the icy pebbles. This process increases $\Sigma_{\rm peb}/\Sigma_{\rm g}$ locally and triggers the streaming instability to form planetesimals at the water-ice line. Furthermore, \cite{Lenz2019} proposed that the dust particles can be trapped at pressure-bumps, which may exist over a wide regions of the disk due to hydrodynamical/magnetic turbulence. Similarly, such places of high enough $\Sigma_{\rm peb}/\Sigma_{\rm g}$ can also facilitate the planetesimal formation. These ideas also motivates our further hypothesis on the birth locations of the protoplanets in the population synethesis study  in \se{pps}.
 
On the other hand, the constant flux ratio assumption also neglects the reduction of $\dot M_{\rm peb}$ due to planet pebble accretion.  
   We give the reason for this simplification as follows. The typical gas disk mass is $1{-}10\%$ of the stellar mass (\fg{diskm}), resulting in the mass ratio between disk pebbles and central host of ${\approx} 10^{-4}{-}10^{-3}$. For comparison, we found that the mass ratio between the pebbles accreted by the planet and the central host ranges from $0$ to maximumly $3 \times 10^{-5}$ (see \fg{mass}). We find that even in the case when the planet reaches the pebble isolation mass of $M_{\rm p}/M_{\star} {\sim} 3\times 10^{-5}$,  the fraction of pebble accreted by the planet is still ${<}10\%$.  Most pebbles in disks are actually accreted to the central hosts rather than accreted by planets in our study.  Our constant flux ratio assumption is therefore justified.

	We present the results of one-millimeter-sized pebbles in the main paper.  In addition, we also consider a case that includes the combined effects of bouncing, fragmentation and radial drift. The corresponding results are summarized in \ap{size}. These two particle size assumptions  nevertheless result in very similar outcomes.   
	
 Based on \Eqs{sigma}{rice} we write the Stokes number of one-millimeter-sized  particles in Epstein regime: 
 \begin{equation}
	 \begin{multlined}
	\taus=  \sqrt{2 \pi} \frac{ R_{\rm peb} \rho_{\bullet}}{ \Sigmag}  \\
	 = \begin{cases}
	 {\displaystyle   0.007
	    \left( \frac{\dot M_{\rm g}}{10^{-9} \Msyr} \right)^{-\frac{1}{3}}  \left(\frac{\rho_{\bullet}}{1.5  \rm \ gcm^{-3} } \right) \left(\frac{R_{\rm peb}}{1 \rm \ mm } \right)  }  \\
\hspace{6.5cm}  [r = r_{\rm ice}],   	    \vspace{0.6cm}    \\ 
	 {\displaystyle  0.2 \left( \frac{\dot M_{\rm g}}{10^{-9} \rm \Msyr} \right)^{-1}
	\left(\frac{M_{\star}}{0.1 \ M_{\odot}} \right)^{-\frac{9}{14}}  \left(\frac{L_{\star}}{0.01 \ L_{\odot}} \right)^{\frac{2}{7}}  } \\
	\hspace{6.cm}   [r = 10 \rm AU],
	\end{cases}
	 \end{multlined}
	\end{equation}
	where $R_{\rm peb}$ and $\rho_{\bullet}$ are the physical radius and internal density of the pebble. The first and second row of the above equation computes the values evaluated at the water-ice line location in the inner viscously heated disk region and at $10$ AU in the stellar irradiation disk region, respectively.

	For fixed-size particles, the Stokes number is inversely proportional to the surface density of disk gas. Since the disk masses decrease both with masses of their central hosts and time,  the Stokes number thus increases with these two quantities. This is clearly shown in  \Fg{St}, which illustrates the evolution of Stokes numbers for millimeter-sized pebbles at the water-ice line (orange) and at $10$ AU (blue) in disks around different mass hosts. For disks around solar-mass stars, no matter where pebbles are, $\taus$ always remains lower than $0.03$. However, for disks around brown dwarfs of  $0.01\Ms$,  $\taus$ approaches unity for pebbles at the ice line when $t{\simeq} 7$ Myr, while for pebbles at $10$ AU, this occurs even at the very early stage when $t {\simeq} 0.5 $ Myr. 
	 
	 The Stokes number determines the efficiency of pebble accretion. When the planet mass is low, the pebble accretion efficiency decreases with $\taus$ in the $3$D regime where the planet accretion radius is smaller than the vertical layer of the pebbles \citep{Morbidelli2015,Ormel2018}.  As the planet grows, it enters the $2$D accretion regime. In this situation, the pebble accretion efficiency increases with $\taus$ \citep{Lambrechts2014,Liu2018}. The above analysis holds as long as the pebbles are marginally coupled to the disk gas, and therefore gas drag is important during the pebble-planet interaction ($10^{-3} {\lesssim} \taus {\lesssim} 1$).  However,  when $\taus$ is much greater than unity, gas drag is negligible during pebble-planet encounters, and pebbles are more aerodynamically like planetesimals. In this case, the actual accretion rate drops substantially \citep{Ormel2010}. Therefore, the preferred Stokes number for pebble accretion ranges from $10^{-3}$ to $1$.
	 
Noticeably, as illustrated in  \fg{diskm} and \fg{St}, we find  that the protoplanets formed at wide orbits around low-mass hosts, especially brown dwarfs, would have difficulty  growing significantly in mass by accreting pebbles. This is because disks around these low-mass objects deplete rapidly in both solids and gas. Furthermore, mm-sized pebbles in disks around such hosts have Stokes numbers larger than unity,  and the corresponding pebble accretion turns out to be very inefficient.

\subsection{Stellar properties}
\label{sec:stellar}
The second modification compared to our previous study is our adoption of the time evolution of central object's luminosity. In \cite{Liu2019b} we assumed that the luminosity of a young star does not evolve during the short disk lifetime (${\lesssim} 10$ Myr), and therefore it only correlates with the stellar mass as $L_{\star} {\propto} M_{\star}^\beta$, where $\beta$ is parameterized to be either $2$ or $1$ for  the pre-main-sequence stars. As an improvement, here we adopt a theoretical calculation of $L_{\star}(M_{\star},t)$ from \cite{Baraffe2003,Baraffe2015}, based on the state-of-the-art evolutionary models, in the (sub)stellar mass range from solar-mass to brown dwarfs below the hydrogen-burning limit. 

The time  and (sub)stellar mass dependences can be seen in \fg{Lstar}. For hosts younger than $10$ Myr, we verify that $L_{\star} {\propto} M_{\star}^{\approx 1-2}$ is still a good approximation. After that, the luminosity tends to follow a cubic or fourth power correlation with the mass of the host. It is worth mentioning that this time-dependent variation is due to the much slower contraction for lower mass stars. For comparison, the contraction of solar-mass stars roughly takes a few tens of million years,  while this can last for $10^{8}$ yr to $10^{9}$ yr  for UCDs.  A complete contour map of $L_{\star}(M_{\star},t)$ is provided in \fg{Ls} in \ap{luminosity}. 
	
Young UCDs are fast rotators \citep{Herbst2007}. We approximate the inner edge of the disk to the  corotation radius of the central host with a typical spin period  $P_{\star}$ of one day, which can be expressed as
\begin{equation}
r_{\rm in} = 0.01 \left(  \frac{M_{\star}}{0.1 \Ms} \right)^{1/3} \left(  \frac{P_{\star}}{1 \ \rm day} \right)^{2/3}  \AU.
\end{equation}

\subsection{Derivation of the characteristic planetesimal mass from streaming instability simulations}
\label{ap:streaming}
	Streaming instability is a powerful mechanism that forms planetesimals by directly collapsing a swarm of mm-sized pebbles \citep{Youdin2005}. Numerous numerical simulations show the robustness of this mechanism for generating planetesimals \citep{Johansen2007,Bai2010,Johansen2012,Simon2016,Schafer2017,Abod2018,Li2019}.  These  planetesimals have a top-heavy mass distribution. The characteristic planetesimal size is roughly $100$ km in the asteroid belt region ($2{-}3$ AU) around a solar-mass star \citep{Johansen2015,Simon2016,Schafer2017,Abod2018}. 

Simulations of  streaming instability are performed in a shearing box, where the code units are typically set as $\rho_{g} {=} H {=}  \Omega_{\rm K}^{-1}{=}1$. The midplane gas density $\rho_{\rm g}$,  the gas disk scale height  $H$ and  the Keplerian angular frequency $\Omega_{\rm K} {= }\sqrt{G M_{\star}/r^3}$ correspond to the units of density, length and time, respectively, where $G$ is the gravitational constant and all these quantities are measured at a radial distance $r$ from the central star.   The unit of mass is given by $\Munit{=}\rho_{\rm g} H^3$.  

 The gas density is set by the dimensionless gravity parameter 
 \begin{equation}
\gamma \equiv  \frac{4 \pi  G \rho_{\rm g}}{ \Omega_{\rm K}^2},
\label{eq:gamma}
 \end{equation}
 where $\gamma$ measures the relative strength between the self-gravity and tidal shear, which depends on the gas density $\rho_{\rm g}$, the stellar mass $M_{\star}$ and the radial distance $r$.
 The mass unit can in turn be written as 
  \begin{equation}
 \Munit =  \rho_{\rm g} H^{3} =  \frac{\gamma h^3}{ 4 \pi }M_{\star},
 \label{eq:M0}
 \end{equation}
 where disk aspect ratio $h{= }H/r $. We note that the variation of $\gamma$ is equivalent to the change of $\rho_{\rm g}$ for fixed $r$ and $M_{\star}$. 
 
In the streaming instability simulations, planetesimal masses are measured in the code unit (in terms of $\Munit$). These masses are expected to follow the same scaling laws when $\Munit$ varies. 
 This means we could extrapolate and obtain the physical mass of the planetesimal when realistic $\rho_{\rm g}$, $M_{\star}$ and $r$  are given.    

 We propose that the characteristic planetesimal mass can be written in a form of  
\begin{equation}
 M_{\rm pl} =  f (\Pi,\gamma, \taus, Z) \Munit.
 \label{eq:Mpl}
 \end{equation}
There are four key physical quantities that regulate the gravitationally collapsing of planetesimals, and therefore sets the control function $f$.   
We give these quantities in a dimensionless manner as follows. 
The radial pressure gradient parameter that measures the sub-Keplerian gas velocity 
\begin{equation}
  \Pi \equiv \frac{\eta v_{\rm K}}{c_{\rm s}} = \frac{\eta}{h},
   \end{equation}
   where $\eta{ =} -h^2 ({\partial {\rm ln P}/  \partial {\rm ln r} }) /2{= } (2-s-q) h^2/2$,  $P$ is the gas pressure, $s$, $q$ are the power law indexes of  the gas surface density and disk aspect ratio, respectively. 
The Stokes number that represents the aerodynamical size of the particles can be written as
\begin{equation}
  \taus \equiv t_{\rm stop} \Omega_{\rm K}^{-1},
     \end{equation}
where $ t_{\rm stop}$ is the particle's stopping time,
   The metallicity  (the surface density ratio between particles and gas)
\begin{equation}
 Z \equiv \frac{\Sigma_{\rm d}}{\Sigma}.
 \end{equation}
 The last one is $\gamma$ that describes the  strength of the self-gravity.

We assume that $f (\Pi,\gamma, \taus, Z)$ is composed of a series of power laws 
\begin{equation}
f (\Pi,\gamma, \taus, Z)  = (Z \gamma)^a \Pi^b \taus^c,
 \end{equation}
 where $a$, $b$, $c$ can be principally obtained from streaming instability simulations.  
 Physically, $Z$ and $\gamma$ (corresponding to the gas density) together represent the disk solid density, a fundamental quantity that sets the collapsing of the pebble clumps and formation of the planetesimals. We expect these two follow the same power law dependence in $f$.    
 \Eq{Mpl} can be rewritten as 
 \begin{equation}
 M_{\rm pl} = C (Z \gamma)^a \Pi^b \taus^c  \Munit. 
 \label{eq:Mpl2}
 \end{equation}
We would like to note that  $M_{\rm pl}{ \propto }\gamma^{a+1}$ since $\Munit$  also linearly scales with  $\gamma$ (\eq{M0}). Importantly, the beforehand condition for the above formula is to trigger filament formation by the streaming instability in the first place. This requires a certain range of disk metallicity, the Stokes number and disk pressure gradient \citep{Carrera2015,Yang2017}.

From Fig. 3 of \cite{Simon2017}, we find that the planetesimal mass barely depend on the Stokes number of the pebbles. 
 We thus drop the $\taus$ dependence on $M_{\rm pl}$ hereafter.  The characteristic mass of the planetesimal can be expressed as
   \begin{equation}
 M_{\rm pl} = C    \left( \frac{ Z}{0.02} \right)^a \left( \frac{ \gamma}{\pi^{-1}} \right)^a \left( \frac{\Pi}{ 0.05} \right)^b   \Munit. 
 \label{eq:Mpl2b}
 \end{equation}
 We note that in the above equation $Z$ represents the local metallicity. As discussed in \se{dust}, this value can differ from global disk metallicity due to different dust concentration mechanisms.

Converting the code unit into the physical unit  based on \eqs{M0}{Mpl2b}, we therefore obtain
    \begin{equation}
     \begin{split}
 \frac{M_{\rm pl}}{M_{\oplus}} = &5 \times 10^{-6}   \left( \frac{C}{5\times 10^{-5}} \right)  \left( \frac{Z}{0.02} \right)^{a}   \left( \frac{\gamma}{\pi^{-1}} \right)^{a+1} \\        
  & \left(\frac{h}{0.05} \right)^{3+b}  
  \left( \frac{M_{\star}}{ 0.1 \ \Ms} \right).
 \label{eq:Mpl3}
 \end{split}
 \end{equation}
  
\begin{figure}[t]
	     \includegraphics[width=\columnwidth]{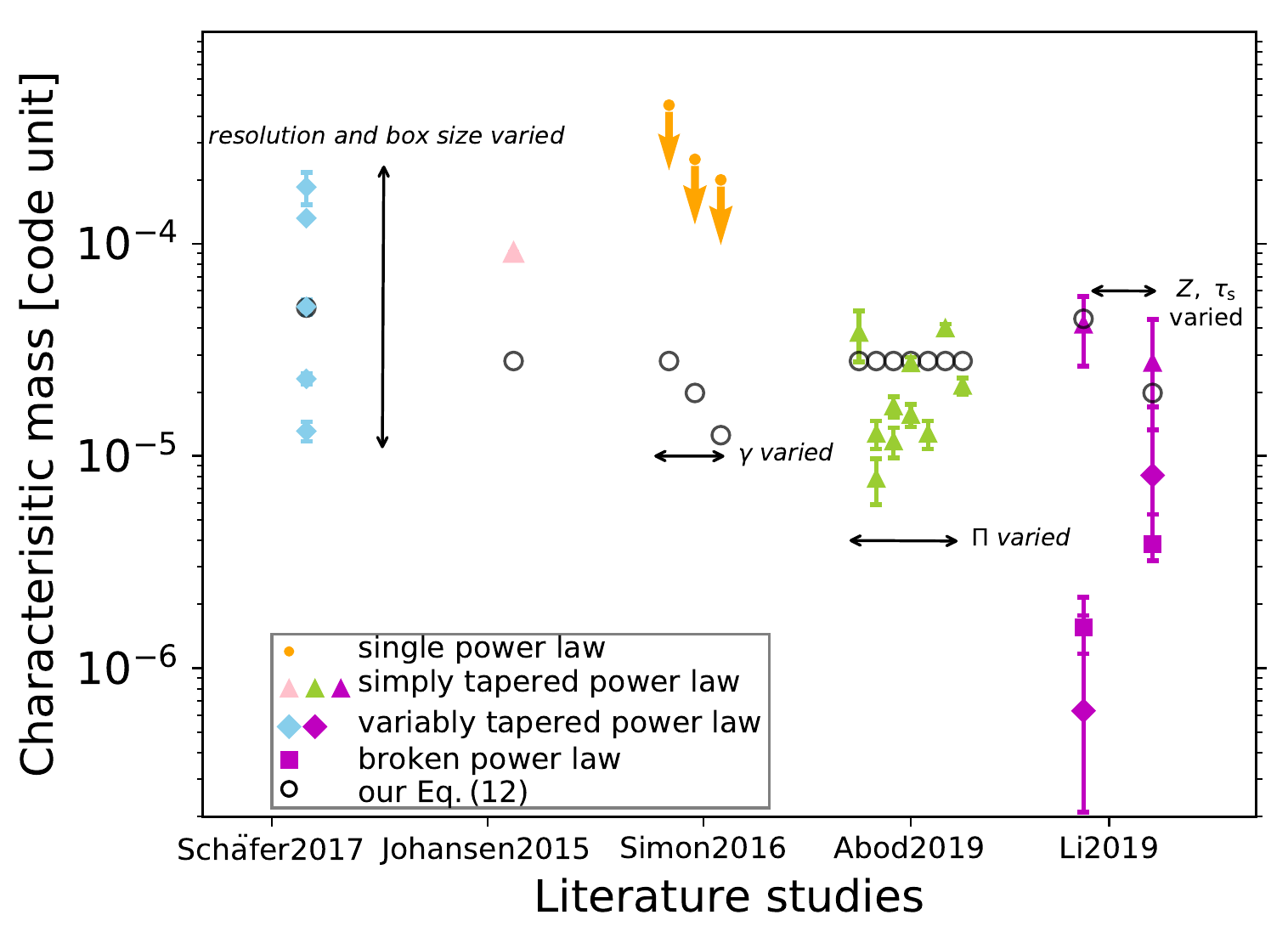}
		\caption{ The characteristic mass comparison among literature studies and our \eq{Mpl2b} calibrated from the results of \cite{Schafer2017}. The planetesimals obtained from streaming instability simulations are fitted by different distributions, including a single power law (dot), a simply tapered power law (triangle), a variably tapered power law (diamond) and a broken power law (square).  The details of these fitting formulas are explained in the main text. The pink, orange,  blue, green and magenta refer to the studies of  \cite{Johansen2015}, \cite{Simon2016}, \cite{Schafer2017}, \cite{Abod2018} and \cite{Li2019}, respectively. We note that the orange dots only refer to the maximum masses of the planetesimals, we add the downside arrows to indicate that the characteristic masses should be lower than these values.  
		 The characteristic masses obtained in each set of parameters are shown in symbols with the same x-axis (see \tb{plt_comparison}).
		 We note that $Z$, $\gamma$, $\Pi$ and $\taus$ are differed by orders of magnitude among these numerical studies.  Our \eq{Mpl2b} agrees reasonably well with the fitting characteristic masses found in these studies.  }
\label{fig:plt_comparison}
\end{figure}
 
There are three unknown parameters $a$, $b$ and $C$ in \eq{Mpl2b}. We explain how we choose these values as follows. Both \cite{Johansen2012} and \cite{Simon2016} explore the role of self-gravity parameter $\gamma$ on the final planetesimal masses.
The inferred power-law index $a$ is between $0$ and $1$ (Fig 13 of \cite{Johansen2012} and Fig. 9 of \cite{Simon2016}). We find that a medium value of $a{=}0.5$ is slightly preferred\footnote{When assuming that the maximum mass and characteristic mass follow the same $\gamma$ dependence, one can compare the declining trend of the orange dots and black circles in \fg{plt_comparison} to obtain a reasonable value of $a$.}.  It results in $M_{\rm pl}$ increases super linearly with $\gamma$ in \eq{Mpl3}. On the other hand, different disk pressure gradients $\Pi$ are explored in \cite{Abod2018}. The inferred $b$ is no larger than $1$ and can be close to $0$ ($b{\sim}0{-}1$, their Fig. 8).  This $\Pi$ dependence on $M_{\rm pl}$ also seems to be non-monotonic. For a conservative choice, we adopt $b{=}0$ here. This leads to $M_{\rm pl} \propto h^3$.
 We choose $C{=}5\times10^{-5}$ based on \cite{Schafer2017}'s table 2. 
 As will be demonstrated later,  our proposed characteristic mass based on the above adopted values exhibits reasonably good agreement with other numerical studies. 

The characteristic mass comparison between the literature studies and our \eq{Mpl2b} is illustrated in \fg{plt_comparison}. In streaming instability simulations, the cumulative mass distribution of the forming  planetesimals is most frequently fitted by a power law plus an exponential decay,  $N_{>}(M) {= } kM^{-\alpha_1}\exp[-(M/M_{\rm exp})^{\beta_1}]$. There are two different approaches for the above fitting. One is the simply tapered power law (STPL), where $k$ and $\alpha_1$ are two free parameters, and $\beta_1$ is fixed to be either $4/3$  \citep{Johansen2015} or $1$   \citep{Abod2018,Li2019}.
The other is the variably tapered power law (VTPW), where $\beta_1$ is an additional free parameter that needs to be fitted \citep{Schafer2017, Li2019}.   Apart from these exponentially tapered power laws, a single power law (SPL) is used in \cite{Simon2016}, and a broken power law (BPL) is explored in \cite{Li2019} (their equation (23)).  The characteristic masses are adopted to $M_{\rm exp}$ in exponentially tapered power laws and $M_{\rm br}$ (turnover mass) in the broken power law. For the single power law fitting, we mark the highest planetesimal mass as an indicator of the upper limit of the characteristic mass. The data used for this comparison is listed in \tb{plt_comparison}.

We find that our proposed characteristic mass in \eq{Mpl2b}  is in good agreement with the literature studies. 
As can be seen in \fg{plt_comparison}, our predicted value underestimates the characteristic mass in \cite{Johansen2015} by a factor of $2{-}3$ (pink triangle). But they are pretty well matched with other studies.  
The above degree of match is impressive, since $\Pi$, $\gamma$, $\taus$ and $Z$ are varied over orders of magnitude among these studies (see \tb{plt_comparison}).  Furthermore,  different fitting formulas and therefore characteristic mass indicators ($M_{\rm exp}$ and $M_{\rm br}$) are adopted in literature. These indicators might differ by a factor of a few from each other even based on the same fitting data (see magenta symbols in \fg{plt_comparison}).  In addition, different planetesimal-finding algorithms among these studies could also induce additional uncertainties in the above fitting masses.
 Overall, here we are confident of the chosen values ($a{=}0.5$, $b{=}0$, $C{=} 5\times10^{-5}$). In future when more numerical simulations are conducted with different disk and pebble parameters, we can optimisze all these parameters in \eq{Mpl2} based on more advanced best-fit algorithms.

\begin{table*}
\begin{threeparttable}
    \centering
    \caption{Characteristic masses in the literature numerical streaming instability studies. The characteristic mass $M_{\rm pl}$ is adopted to be $M_{\rm exp}$ from the STPL and VTPL fittings, or to be $M_{\rm br}$  from the BPL fitting (see Eq.(23) of \citealt{Li2019}). For the SPL fitting, we adopt the largest planetesimal mass as an upper limit of $M_{\rm pl}$.  All these masses are provided in the code unit ($\Munit$).  The data is adopted from \cite{Johansen2015},  \cite{Simon2016}, \cite{Schafer2017}, \cite{Abod2018} and \cite{Li2019}. }
    \begin{tabular}{lllllllllllll|l|}
        \hline
        \hline
    Reference & $Z$ &  $\gamma$  & $\Pi$ &$\taus$   & fitting type & characteristic mass   [$ \Munit$]  \\
        \hline
  \cite{Johansen2015} & 0.02 &   0.1 & 0.05 & 0.3 & STPL & $9.1\times10^{-5}$         \\ \hline
  \multirow{3}{*}{\cite{Simon2016}} & 0.02 &  0.02 & 0.05 & 0.3 & SPL  & $4.5 \times 10^{-4}$ \tnote{a} \\
  & 0.02 &  0.05 & 0.05 & 0.3 & SPL  &  $2.4\times 10^{-4}$ \\
   & 0.02 &  0.05 & 0.05 & 0.3 & SPL  &  $2.0 \times 10^{-4}$ \\ \hline
 \cite{Schafer2017} & 0.02 &  $1/\pi$ & 0.05 & 0.314 & VTPL  & their table 2 \tnote{b} \\ \hline 
 \cite{Abod2018} & 0.1 &   0.02 & 0-0.1 & 0.05 &    STPL &   their table 1 \tnote{c}  \\ \hline
  \multirow{2}{*}{\cite{Li2019}} & 0.1  &   0.05 & 0.05 & 2.0 &     STPL, VTPL, BPL & $4.1 \times 10^{-5}, 6.3\times 10^{-7}$,$1.6\times 10^{-6}$   \\ 
  & 0.02 &   0.05 & 0.05 & 0.3 &     STPL, VTPL, BPL & $2.7 \times 10^{-5}, 8.1\times 10^{-6}$, $3.8 \times 10^{-6}$   \\ 
        \hline
        \end{tabular}
      \begin{tablenotes}
       \footnotesize
       \item[a]  the mass of the largest planetesimal (the upper limit of $M_{\rm pl}$) 
        \item[b] There are six simulations with different numerical resolutions and box sizes.  
         \item[c] There are ten simulations with different $\Pi$ and the time when particles' self-gravity is initiated.  
    \end{tablenotes}
     \end{threeparttable}
    \label{tab:plt_comparison}
\end{table*}

	\subsection{Initial mass of the protoplanets}
	\label{sec:embryo}
	
	\begin{figure*}[t]
	    \includegraphics[width=\columnwidth]{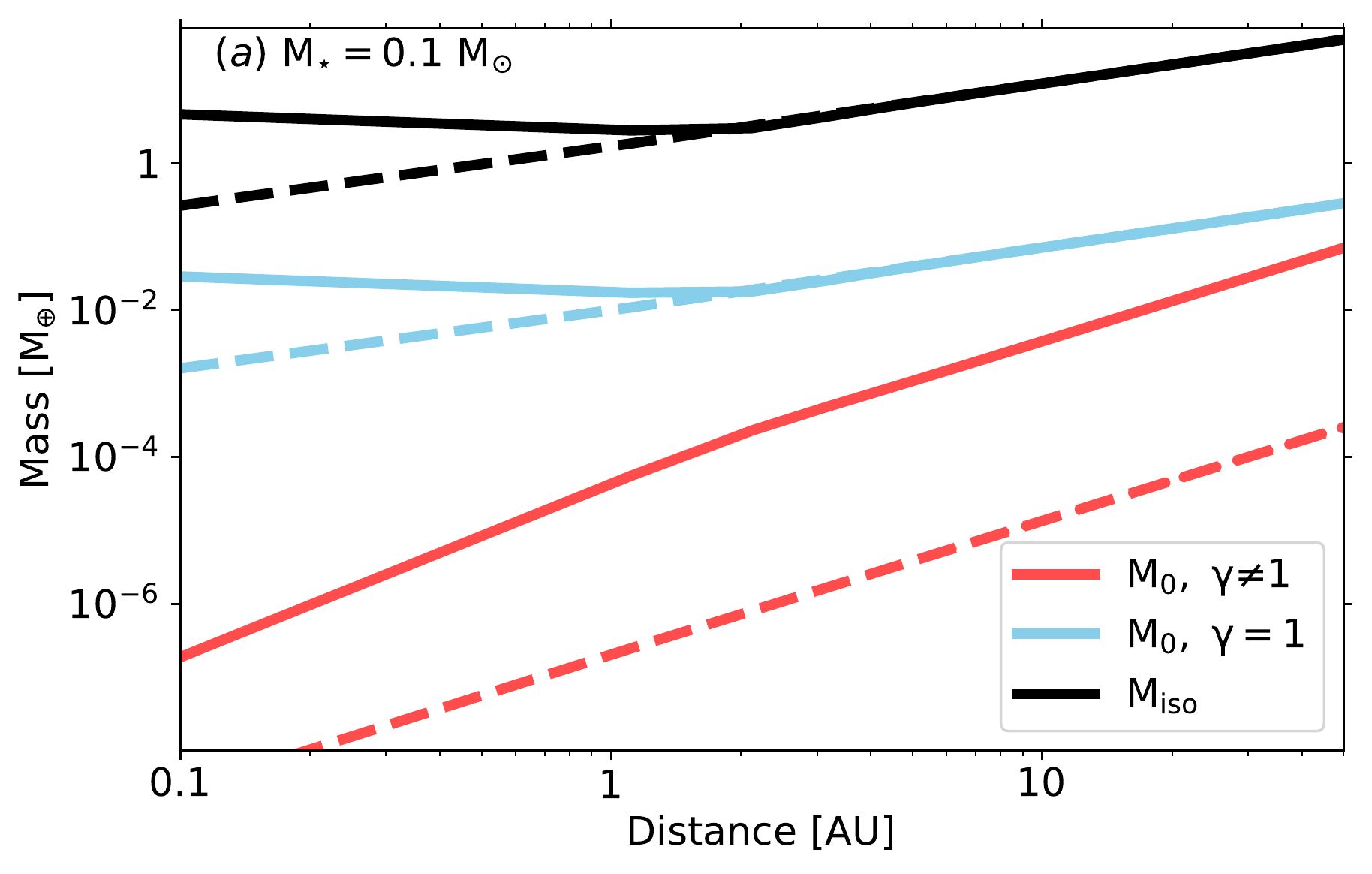}
	   \includegraphics[width=\columnwidth]{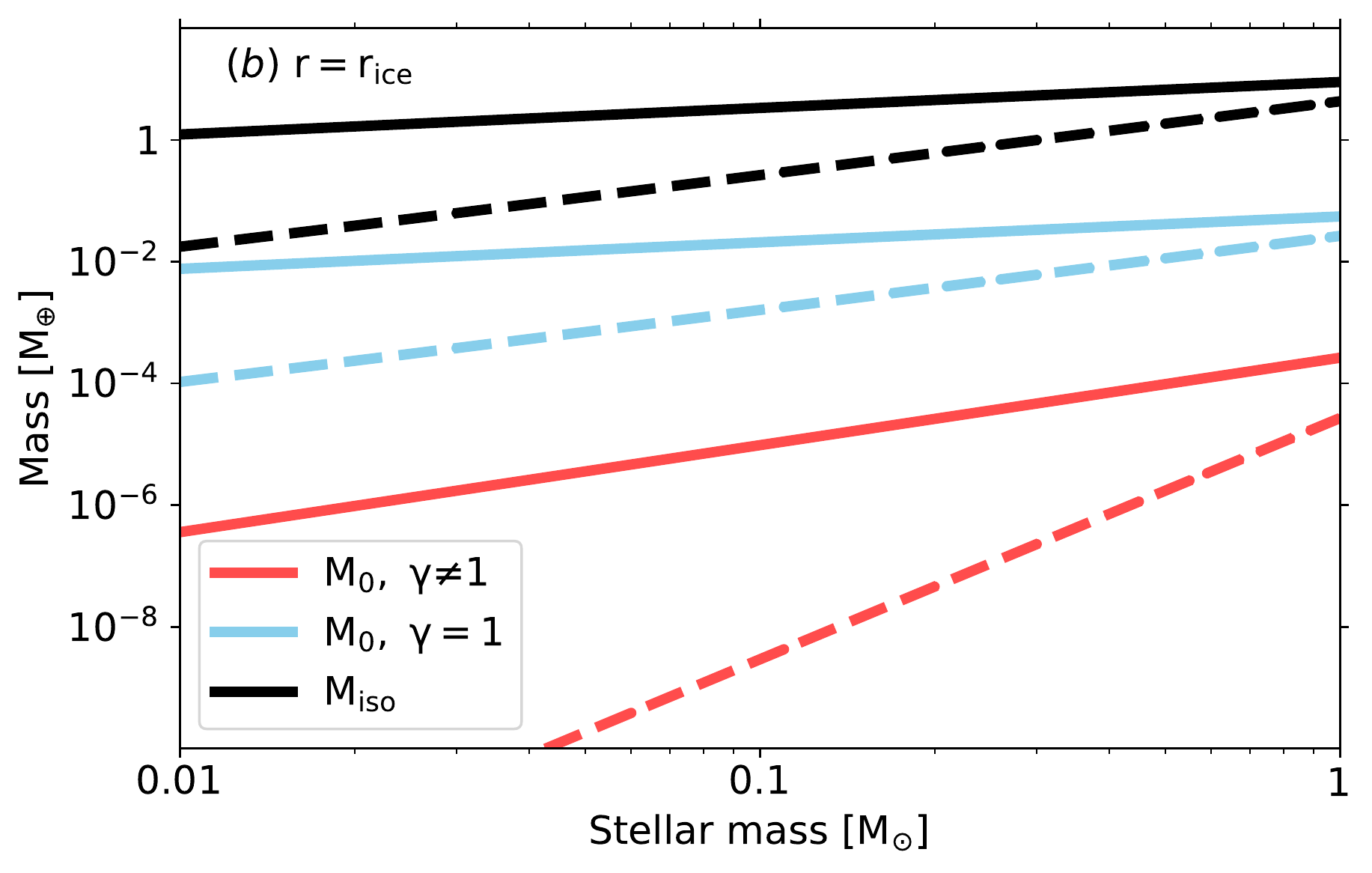}
		\caption{  Mass of protoplanet $M_{0}$ (\eq{Mem_new}) formed by streaming instability as a function of radial distance (left) and host's mass (right).  The cyan and red lines represent the protoplanets in self-gravitating disks ($\gamma {=}1$) and  non-self-gravitating disks ($\gamma {\ne}1$), respectively, whereas the black line corresponds to the pebble isolation mass.  The birth time of the protoplanet is assumed at $t=0$ Myr (solid) yr or $3$ Myr (dashed). The mass of central host is $ 0.1 \Ms$ in  panel (a) and the birth location of the protoplanet is at $r_{\rm ice}$ in panel (b).  When protoplanets form by streaming instability, their birth masses increase with both the masses of central hosts and radial distances. }
	\label{fig:plt}
	\end{figure*}

Based on streaming instability simulations (\eg , \citealt{Schafer2017}), the mass of the largest body $M_0$ is typically about two to three orders of magnitude higher than the characteristic mass (see Fig.7 in \cite{Liu2019a}). 
 As demonstrated in  \cite{Liu2019a}, this most massive planetesimal would dominate the following mass growth and dynamical evolution of the whole population.
 In this population synthesis study, we adopt the largest planetesimal formed by streaming instability as our starting point, and we call it the protoplanet hereafter. The subsequent growth and migration of one single  protoplanet is investigated in this study. 

Assuming that the disk solids would be locally enriched to satisfy the $Z$ criterion, we drop the $Z$ dependence in \eq{Mpl3} hereafter. The mass of the initial protoplanet is given by   
    \begin{equation}     
    \begin{split}
\frac{ M_0}{M_{\oplus}} = & 2 \times 10^{-3}  \left( \frac{f_{\rm plt}}{400} \right) \left( \frac{C}{5\times 10^{-5}} \right)   \left( \frac{\gamma}{\pi^{-1}} \right)^{a+1} \\
&  \left(\frac{h}{0.05} \right)^{3+b}  \left( \frac{M_{\star}}{ 0.1 \ \Ms} \right),
 \label{eq:Mem_new}
  \end{split}
 \end{equation}
where $f_{\rm plt}$ is the ratio between the maximum mass and the characteristic mass. In this study, we set $f_{\rm plt}{=}400$, $a{=}0.5$, $b{=}0$ and $C{=} 5\times 10^{-5}$ to be the fiducial values.  
	
 Since $\gamma $ is a disk-related parameter, the key issue is to understand how $\gamma$ scales with $M_{\star}$ and $r$ under different disk conditions. Two types of disks (or two disk evolutional phases) are explored, self-gravitating and non-self-gravitating,  which result in $\gamma$ being either equal to $1$ or much less than $1$.

	First, we consider the case when the disk is everywhere gravitationally unstable (Toomre Q ${\approx}1$). This requires a young and massive disk. 
	In this situation $\gamma{=}1$, and the mass of the forming protoplanet is independent of $\gamma$. Therefore, we do not need to know the gas density $\rho_{\rm g}$ directly, since its value is encoded in $\gamma$. As can be seen from \eq{Mem_new} the protoplanet's mass exhibits a similar scaling as the pebble isolation mass, also proportional to $ h^3 M_{\star}$ \citep{Lambrechts2014b,Bitsch2018}. 
	
	On the other hand, during a later phase when the disk is non-self-gravitating, $\gamma {\ne}1$ and the actual value depends on the gas surface density.  We then calculate $\gamma$ based on our two-component disk model with an inner viscously heated region and an outer stellar irradiated region (see \ap{diskmodel}).  In a steady state viscous accretion disk, the gas surface density is related to the mass accretion rate and viscosity such that $\dot M_{\rm g}{ =} 3 \pi \Sigma_{\rm g} \nu$, where  $\nu {= } \alphag c_{\rm s}^2/\Omegak$, $\alphag$ represents the angular momentum transfer  coefficient and $c_{\rm s}$ is the gas sound speed. 
	Based on the definition of $\gamma$, we derive that  $\gamma \propto \dot M_{\rm g}/\alphag c_{\rm s}^3 $ in this case.  It means that by given disk temperature and $\alphag$, $\gamma$ is proportional to $\dot M_{\rm g}$.

	 For the non-self-gravitating disk, after the protoplanets form, the disk evolution follows the nominal two-component model, where the disk surface density and aspect ratio are adopted from \eqs{sigma}{aspect} and the time-evolution is based on \eq{mdot}. However, how $ \rho_{\rm g}$ evolves in a self-gravitating disk  is a more subtle issue, which relies on additional assumptions. 
	 In principle,  $ \rho_{\rm g}$ is higher in a self-gravitating disk.  Even though a young disk is self-gravitating at early times, its density decreases and the disk would finally evolve into the non-self-gravitating state. 
	Here we assume that once protoplanets form in a self-gravitating disk, the density quickly adjusts to that of the non-self-gravitating disk.  This means that $\rho_{\rm g}$ is the same for these two types of disks after the protoplanet formation. Then the key difference would be the birth masses of protoplanets.  Such a  treatment is conservative for the self-gravitating disk case.  However, the advantage is that the model is reasonably simplified and no additional parameters are needed to be included.    


	 \Fg{plt}a shows the mass of the protoplanet $M_0$ formed by streaming instability as a function of the disk location either in a self-gravitating disk ($\gamma{=}1$, cyan) or in a non-self-gravitating disk ($\gamma{\ne} 1$, red) around a $0.1 \Ms$ star.  The solid and dashed lines correspond to the formation time of the protoplanet at $t{=}0$ Myr and $3$ Myr, respectively.  The black line represents the pebble isolation mass for comparison. We can see clearly that the initial protoplanet has the same mass scaling as the pebble isolation mass in a self-gravitating disk. When the protoplanet forms early (solid cyan), at the inner sub-AU region, the disk is viscously heated and the aspect ratio is almost independent of the distance (flat disk, \eq{aspect}), therefore $M_0$ and $M_{\rm iso}$ are insensitive to the radial distance.  At the region further out where the disk is heated by stellar irradiation, the aspect ratio increases as $r^{2/7} $ (flaring disk), and the protoplanet's mass is nearly proportional  to $r$. In a non-self-gravitating disk ($\gamma{\ne}1$), $M_0$  additionally depends on $\gamma$, which also increases with $r$. This simply reflects the fact that the disk self-gravity is likely to overwhelm the tidal shear as the radial distance increases.  As a result, we find that the mass of the protoplanet increases super-linearly with the distance. 
	 
	We also clearly see how the mass of the protoplanet formed by streaming instability varies with time. When the protoplanet forms late at $3$ Myr (dashed), the disk is entirely heated by stellar irradiation, and as mentioned the protoplanet's mass is purely proportional to $r$ in the self-gravitating disk ($\gamma{=}1$).  Since $\gamma$ decreases with time in the non-self-gravitating disk,  the late forming protoplanet is less massive compared to the one that forms early.  
	 	 
	 We also plot the mass of the protoplanet as a function of the stellar mass for the above two disks at the water-ice line in \fg{plt}b. We find that the protoplanet's mass increases weakly with the stellar mass in the self-gravitating disk, whereas this stellar mass dependence is more pronounced in the non-self-gravitating disk. We explain the reasons as follows.  First, based on \eqs{Mem_new}{hice} $M_0(\gamma{=}1)$ is proportional to $\dot M_{\rm g}^{2/3}$ in the early phase when $r_{\rm ice}$ resides in the inner viscously heated region. Second, as shown in \fg{diskm}a, the difference of disk accretion rates among different mass stars at the early phase is relatively small, implying a weak $M_{\star}$ dependence on $\dot M_{\rm g}$. Combined with the above two factors, at early time the protoplanet's mass modestly increases with the stellar mass in the self-gravitating disk (solid cyan).  Nevertheless, in the non-self-gravitating disk, $M_0(\gamma {\ne}1)$ additionally correlates with $\gamma$.  
When taking that into account, we find that $M_0$ exhibits a stronger $M_{\star}$ dependence in a non-self-gravitating disk than in a self-gravitating disk. 
	 
As discussed in \se{disk}, $\dot M_{\rm g}$ is roughly proportional to $M_{\star}$ at early times, and the $\dot M_{\rm g}{-}M_{\star}$ correlation becomes steeper as disks evolves. We therefore expect that the mass of the late forming protoplanet  has a steeper stellar mass dependence than the one forms early. We find in \fg{plt}b that the late forming protoplanet in a non-self-gravitating disk around a low-mass host is very small. For instance, $M_0(\gamma{\ne} 1) \simeq 10^{-9} \Me$ when it forms at $t{=}3$ Myr around a $0.1 \Ms$ star, corresponding to $10$ km in size.  
	
In summary, the mass of the protoplanet generated by streaming instability correlates with the mass of the central host and the disk location. For the same (sub)stellar host, low-mass protoplanets form at close-in orbits while massive ones form at further distances. For the same disk location, low-mass protoplanets form around less massive brown dwarfs while high mass protoplanets form around higher mass stars. The above dependencies are more pronounced  for the protoplanets that form in  late non-self-gravitating disks than those in early self-gravitating disks.

\section{Growth track of single protoplanet}
	\label{sec:growth}
The detailed treatment of planet growth and migration can be found in Sections 2.2 and 2.3 of \cite{Liu2019b}. Here we provide an analytical comparison between the migration and growth of a growing protoplanet around different mass hosts. 

Protoplanets and low-mass planet undergo type I migration \citep{Kley2012}, and the corresponding migration rate is 
	 \begin{equation}
	  \begin{aligned}
	 \frac{{\rm d} r}{{\rm d}t } & = - k_{\rm mig}\left( \frac{\Mp}{M_{\star}} \right)  \left( \frac{\Sigmag r^2}{M_{\star}} \right) \left( \frac{1}{h} \right)^2 V_{\rm K} \\
	 & = -\frac{k_{\rm mig}}{3 \pi \alphag} \left( \frac{\Mp}{M_{\star}} \right)  \left( \frac{\dot M_{\rm g}}{M_{\star}} \right) \left( \frac{1}{h} \right)^4 r.  
	  \label{eq:migration}
	   \end{aligned}
	\end{equation}
	where $k_{\rm mig}$ is the migration coefficient and the latter equation is derived from the steady-state disk assumption where $\dot M_{\rm g} = 3 \pi \nu \Sigmag$. It should be noted that we also use two different $\alpha$ parameters in this work, where the global disk viscous angular momentum transfer coefficient $\alphag{=}10^{-2}$, and the level of the local disk turbulence $\alphat{=}10^{-4}$. By a given disk accretion rate,  $\alphag$ determines the gas disk surface density, and thus planet migration. On the other hand, planet growth is set by $\alphat$ which corresponds to the the pebble settling and pebble accretion efficiency (see Section 2 of \cite{Liu2019b}). 
	
	The mass accretion rate of a planet in the $2$D shear-dominated pebble accretion regime can be found in \cite{Liu2018}  as 
	\begin{equation}
	 \frac{{\rm d}\Mp}{{\rm d}t } = \varepsilon_{\rm PA} \dot M_{\rm peb} = k_{\rm PA} \left( \frac{\Mp}{M_{\star}} \right)^{2/3}  \left( \frac{\dot M_{\rm peb}}{ \eta \taus^{1/3} } \right), 
	 \label{eq:growth}
	\end{equation}
	where $\varepsilon_{\rm PA}$ is the pebble accretion efficiency, $k_{\rm PA} =0.24$ is the numerical prefactor in shear-dominated  regime.

Based on  \Eqs{migration}{growth}  we can obtain the differential equation for the growth track $M_{\rm p}(r)$ as \citep{Lambrechts2014,Johansen2018} 
	 \begin{equation}
	 \frac{{\rm d} \Mp}{{\rm d} r} =\frac{ 6 \pi \alphag  h^2  \xi}{ (2-s-q)\taus^{1/3}} \left( \frac{k_{\rm PA}}{  k_{\rm mig}} \right) \left( \frac{\Mp}{M_{\star}}  \right)^{-\frac{1}{3}}      \left( \frac{ M_{\star} }{r} \right), 
	\end{equation}
	where $s$ and $q$ are power-law indexes of disk surface density and aspect ratio.  
	The solution of the above equation is expressed as  
	 \begin{equation}
	\Mp^{4/3} - M_{\rm p0}^{4/3}= -K (r^{2q} - r_0^{2q}), 
	\label{eq:track} 
	\end{equation}
	where $K{=}8\pi \alphag \xi  h_{\rm AU}^2M_{\star}^{2/3}  \taus^{-1/3} k_{\rm pA}/k_{\rm mig}$ and the subscript $0$ means the initial value, $h_{\rm AU}$ refers to the disk aspect ratio at $1$ AU. A larger $K$ indicates the growth is more pronounced than the migration, and the growth track stars to turn over at a higher planet mass (\fg{growth}). In the limit of  $M_{\rm p} {\gg} M_{\rm p0}$ and $r{=}0$, $\Mp {=}(K r_0^{2q})^{3/4}$.  We  can see that the final mass of the planet depends on both $K$ (correlated with the disk metallicity $\xi$, the stellar mass $M_{\star}$ and the particles' Stokes number $\taus$) and the initial position $r_0$.

	\begin{figure}[t]
	     \includegraphics[scale=0.55, angle=0]{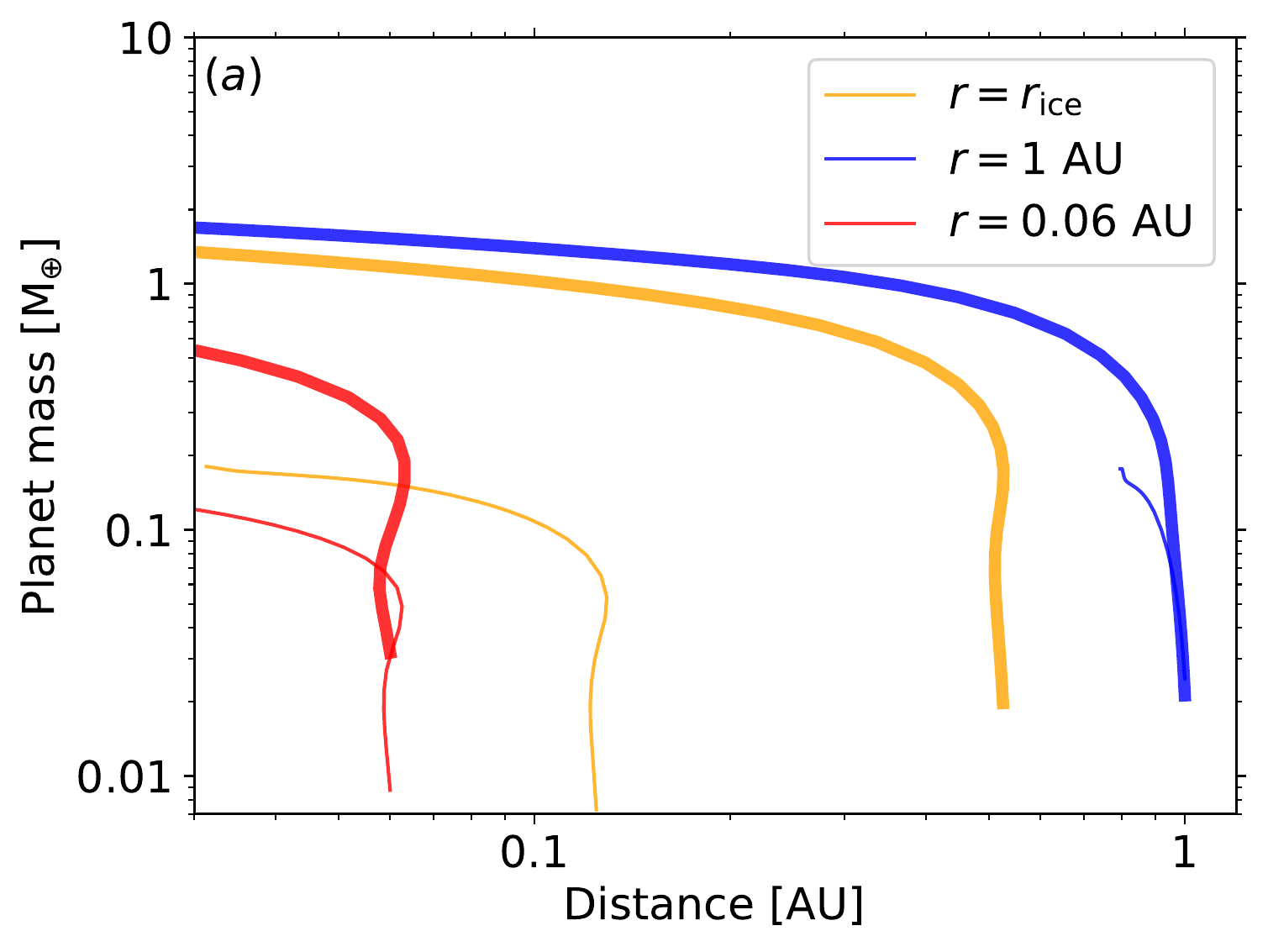}
	      \includegraphics[scale=0.55, angle=0]{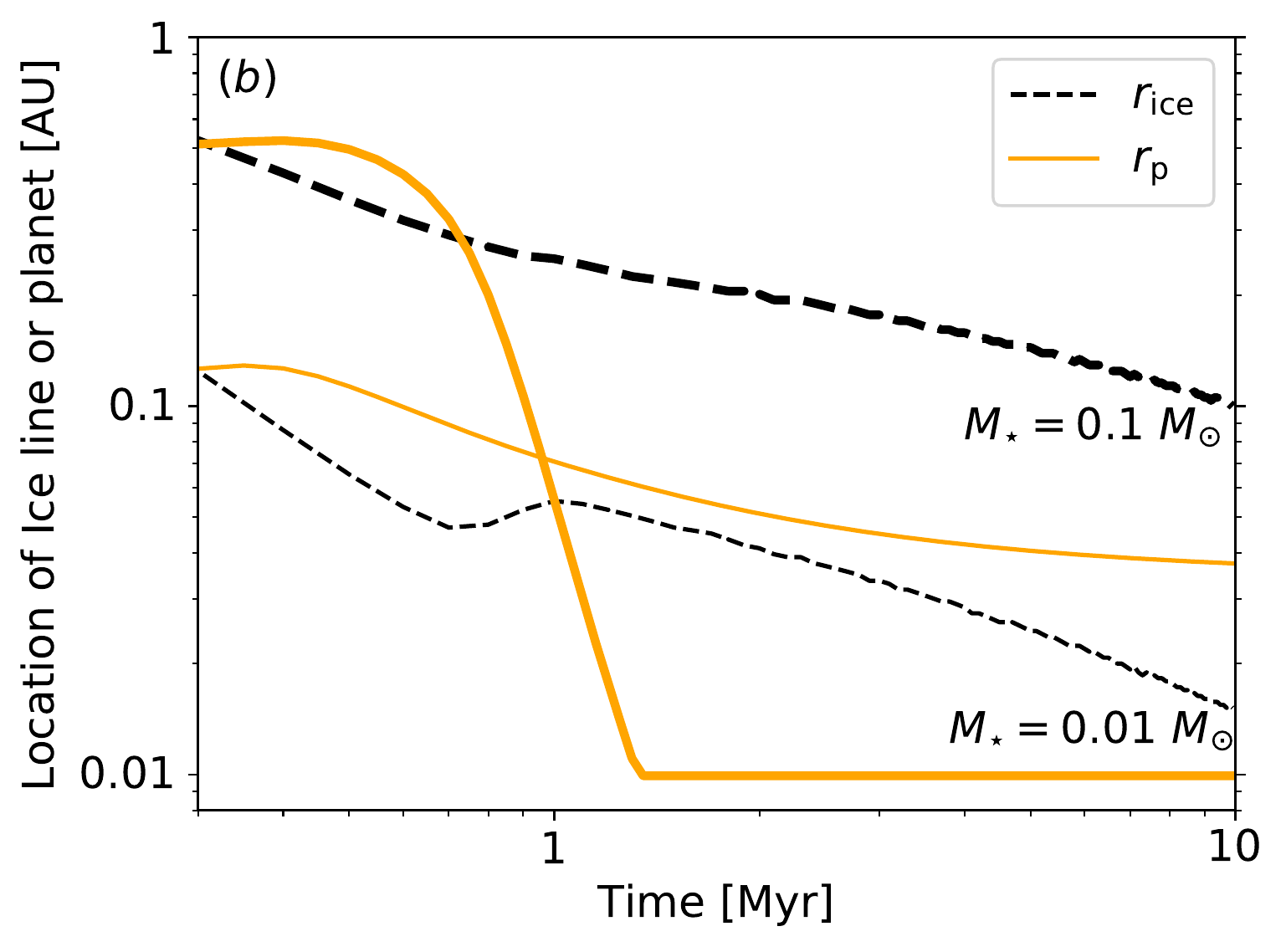}
		\caption{ (a): Growth tracks of protoplanets around a $0.01\Ms$ brown dwarf (thin) and a $0.1\Ms$ star (thick). The birth locations of protoplanets are at $r{=} 0.06 \AU$ (red), the water-ice line (orange), and $r{= }1 \AU$ (blue).  The protoplanets are all assumed to form at $t=0$ Myr and their masses are adopted from \eq{Mem_new} with $\gamma{=}1$. In the beginning, disks are dominated by viscous heating, and $r_{\rm ice} {=}0.12$ AU and $0.55$ AU for systems around the  $0.01 \Ms$ brown dwarf and $0.1 \Ms$ star, respectively.   (b) Time evolution of the water-ice line locations (black) and planet's semi-major axis (orange) for systems around different mass stars. In these two panels  the pebble-to-gas mass flux ratio is adopted to be $3\%$, corresponding to the stellar metallicity $ [\rm Fe/H]$ of $0.48$. The inward movement of the ice line is faster in disks around brown dwarfs, while the migration of the protoplanets formed at the ice line is more significant around low-mass M-dwarfs. The water fraction of planets formed at the water-ice line increases with the stellar mass. 
		}
	\label{fig:growth}
	\end{figure}

 \Fg{growth}a shows  the growth track for the protoplanets at different disk locations around a $0.01 \Ms$  brown dwarf (thin line) and a $0.1 \Ms$ star (thick line). We adopt the masses of protoplanets from \eq{Mem_new} for the self-gravitating disk ($\gamma{=} 1$), and assume that in all cases the protoplanets form at $t{=}0$ yr. Based on \eqs{mdot}{rice}, initially water-ice lines reside inside the viscously heated regions that $r_{\rm ice} {=} 0.12 \AU$ and $0.55 \AU$ for disks around the corresponding brown dwarf and low-mass star, respectively. We find in \fg{growth}a that the turn-over mass when the migration overwhelms the growth is higher for planets around more massive stars. For the case of $r_0 {= }r_{\rm ice}$ (orange), the protoplanet around the brown dwarf undergoes substantial migration when $\Mp {\gtrsim} 0.1 \Me$, whereas it starts to migrate at $0.5 \Me$ around the $0.1\Ms$ star. The planet also reaches a higher mass when it orbits around the low-mass star than the brown dwarf.  As mentioned, both the maximum mass and turn-over point in the growth track are set by $K$ (\eq{track}), which increases with the stellar mass. A similar stellar mass dependence can also be found in \fg{growth}a for the cases of $r_0{<}r_{\rm ice}$ (red) and $r_0{>}r_{\rm ice}$ (blue). 
	
	Noticeably, the final planet mass strongly depends on the birth location.  For systems around $0.1 \Ms$ stars, the protoplanets can grow up to $2 \Me$ and $0.5 \Me$ when their birth locations are $ 1 \AU$ and $0.06 \AU$. For systems around brown dwarfs, we can see that the growth is insignificant when the protoplanets form at $1 \AU$.  This is because the disks around brown dwarfs deplete rapidly, and therefore Stokes number of the mm-sized particles becomes very high, resulting in inefficient pebble accretion.    Hence, for systems around brown dwarfs,  larger planets can only form when their birth locations are close to water-ice lines ($r_0 {\sim} r_{\rm ice}$).    
	
	Furthermore, we also plot the evolution of the ice line (black) and planet (orange) in low-mass star and brown dwarf disks in \fg{growth}b. We find that ice line moves inward much quicker around brown dwarfs, due to a faster depletion of disk gas. On the other hand, the protoplanets formed at the ice line grow less around brown dwarfs, and thus their migration is slower. We can clearly see this from an illustrated case in \fg{growth}b.  As a result, more protoplanets in brown dwarf disks are likely to stay outside of the ice line and contain higher water fractions compared to those around low-mass stars.   This trend is also observed in planet population synthesis simulations (\fg{mass}a), which will be further discussed in \se{pps}.

\section{Population synthesis study}
\label{sec:pps}
In this section we perform Monte Carlo simulations to study the growth and migration of a large sample of protoplanets. The initial conditions for simulated protoplanets are described in \se{IC}, where the distributions of  model parameters are given in \tb{parameter}. We show the  properties (masses, semimajor axes and water fractions) of the forming planets in the Monte Carlo sampling plots in \se{MC}.

\subsection{Initial condition setup}
\label{sec:IC}

Our adopted disk accretion rate follows a lognormal distribution, with the mean value given by \eq{mdot} and a standard deviation $\sigma$ of $0.3$.  The dispersion among the measured disk accretion rates onto certain spectral type stars can be very large, ranging from $0.4$ dex in Lupus \citep{Alcala2014} to $1$ dex in Chamaeleon I star forming regions  \citep{Manara2016,Mulders2017}. We note that for individual stars, the disk accretion rates also vary with time; they are high in early and low in late phase (Fig. 12 in \cite{Manara2012}).  
The age difference among individual objects can be a few Myr even in the same star forming cluster. Therefore, the large dispersion in the accretion rate could be partially caused by this age difference. Since we already account for the time-evolution of $\dot M_{\rm g}$, the standard deviation considered here should be smaller than the observed values. As the same in \cite{Liu2019b},  we choose the $0.3$ dex standard deviation.  

 Based on the scaling analysis from the streaming instability simulations, the masses of the protoplanets are adopted from \eq{Mem_new}. The disks are assumed to be either self-gravitating ($\gamma{=}1$) or non-self-gravitating ($\gamma{\ne}1$). The starting positions of these bodies are either only at the water-ice lines (Scenario A), or log-uniformly distributed from $0.01$ AU to $10$ AU (Scenario B). The formation time of these protoplanets are uniformly distributed from $t {=}0.1$ Myr to $0.5$ Myr.  The masses of the central hosts $M_{\star}$ are log-uniformly adopted from the brown dwarfs of $0.01 \Ms$ to the low-mass stars of $0.1 \Ms$.  The stellar metallicities $Z_{\star}$, corresponding to the disk pebble-to-gas mass flux ratios, range uniformly from $-0.5$ to $0.5$. As proposed in \se{disk}, the pebbles are assumed to be all one-millimeter in size. Simulations are terminated at $t{=}10$ Myr, which is the typical lifetime of gaseous disks around these ultra-cool dwarfs. We also perform simulations with $t{=}20$ Myr and find that the final planet mass difference between two cases is less than a few per cent.

\begin{table}
    \centering
    \caption{Adopted parameter distributions for the population synthesis study in \se{pps}}
    \begin{tabular}{lclclclclclc|}
        \hline
        \hline
        Parameter   &   Description  &   \\ 
        disk model   &  viscously heated + stellar irradiation \\ 
         $\dot M_{\rm g0}$  [$\Msyr$] & $ 10^{\mathcal {N}(\mu, {\sigma}^{2} ) }$, $\mu$ is from \eq{mdot} and $\sigma=0.3$ \\ 
         $ M_{\rm p0} [\Me] $ &  \eq{Mem_new}, self-gravitating disk ($\gamma {= }1$)\\
           &  \ \   non-self-gravitating disk ($\gamma {\ne} 1$)   \\
                   $ r_0$ [AU] &  Scenario A: the ice line  \\
           &  Scenario B:  $\rm{\log U}(0.01,10)$ \\
           $ t_0$ [Myr] & $\rm{U}(0.1,0.5)$ \\ 
            $ M_{\star}$ [$\Ms$]  & $\rm{\log U}(0.01,0.1)$  \\ 
            $Z_{\star} $ &  $ \rm{U}(-0.5,0.5)$ \\ 
            pebble size &  one millimeter \\ 
      \hline
        \hline
    \end{tabular}
    \label{tab:parameter}
\end{table}

\begin{figure*}[tbh]
 \includegraphics[scale=0.8, angle=0]{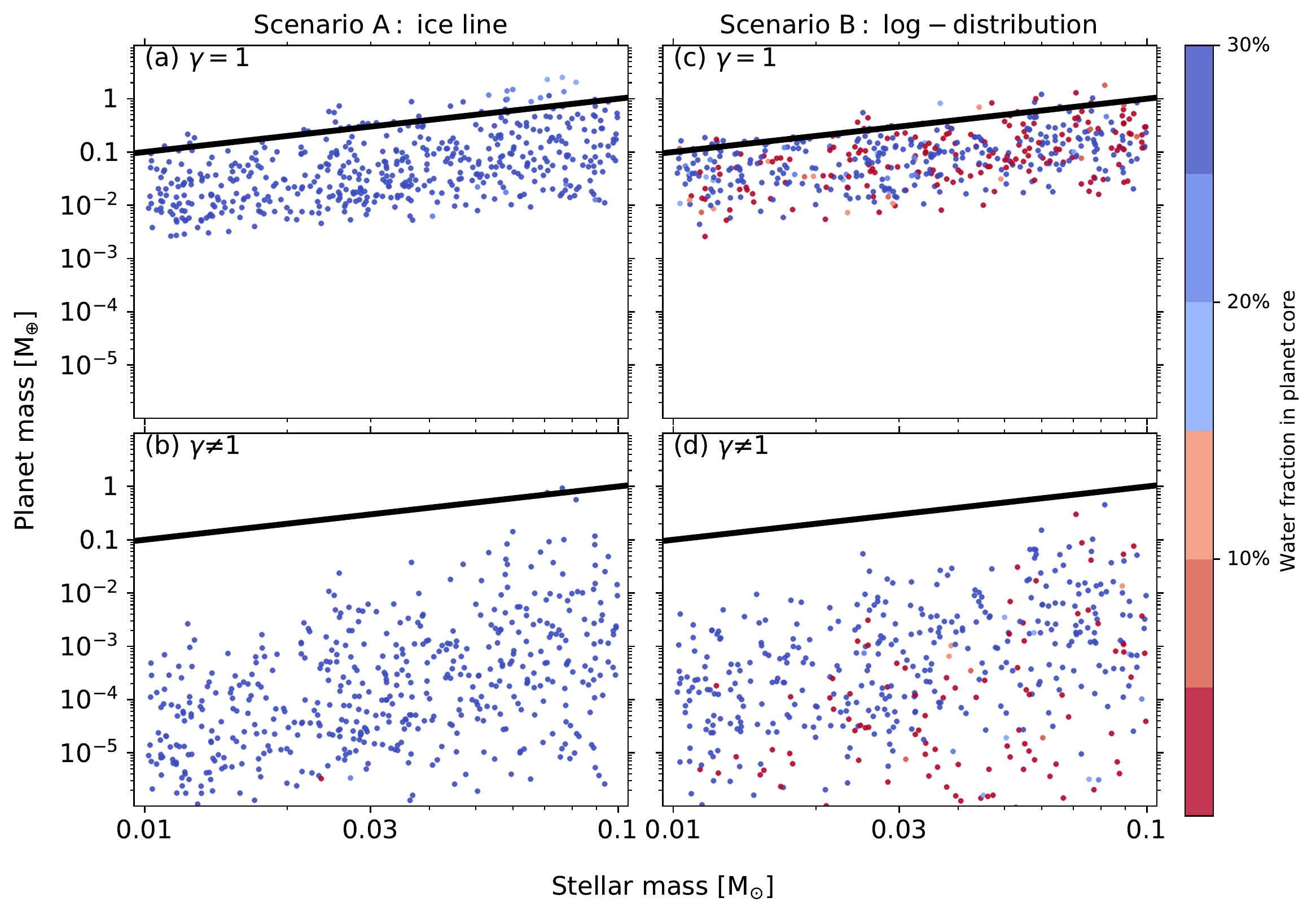}
       \caption{  
 Monte Carlo sampling plot of the planet mass as a function of the stellar mass, with the ice line planet formation model (Scenario A) in the left, the log-uniform distributed planet formation model (Scenario B) in the right, the self-gravitating disks ($\gamma{=}1$) in the top and the non-self-gravitating disks ($\gamma {\ne} 1$) in the bottom.  The color corresponds to the  water mass fraction in the planetary core.  The black line stands for a constant planet-to-star mass ratio of $3 {\times}10^{-5}$ motivated by \cite{Pascucci2018}, which also approximates to the pebble isolation mass scaling \citep{Lambrechts2014b,Liu2019b}. There is still an obvious $M_{\rm p}{-}M_{\star}$  correlation for the central hosts in the ultra-cool dwarf mass regime of $0.01 \Ms$ and $0.1 \Ms$. This correlation is independent of the birth locations of protoplanets but relies on the disk conditions. Protoplanets can grow to Mars-mass around $10 M_{\rm Jup}$ brown dwarfs when they are born in the self-gravitating disk phase,  while they only reach $10^{-3}{-}10^{-2} \Me$ when the the protoplanets form later in the non-self-gravitating disk phase. Planets with ${\gtrsim} 15\%$ water can form at the water-ice line, while protoplanets formed over a wide range of disk distances end up with a distinctive, bimodal water mass.
    }
\label{fig:mass}
\end{figure*} 

\subsection{Monto Carlo simulations}
\label{sec:MC}

The resulting planet distribution obtained from our Monto Carlo simulation is illustrated in \fg{mass}, which shows planet mass versus stellar mass. The masses of initial protoplanets are either obtained from the assumed self-gravitating disks  ($\gamma{=}1$, top), or from the non-self-gravitating, nominal two-component disks ($\gamma {\ne}1$, bottom), and their birth locations are either at the water-ice lines (left) or log-uniformly distributed over the entire disk regions (right). The color corresponds to the planet water mass fraction, and the black line refers to a constant planet-to-star mass ratio ($M_{\rm p}/M_{\star}$) of $3 {\times} 10^{-5}$.  Based on the Kepler data, \cite{Pascucci2018} find that the most abundant planet population around GK and early M stars has such a universal planet-to-star mass ratio. Extrapolating this characteristic planet mass to the stellar mass regime considered in this work, we obtain a Mars-mass planet around a  $10 \  M_{\rm Jup}$ brown dwarf and an Earth-mass planet around a $0.1\Ms$ star. 

We note that the planet mass growth is limited by the black line shown in \fg{mass}. Such a characteristic planet mass derived from \cite{Pascucci2018} matches approximately well with the pebble isolation mass, where the planet at this mass stops the inward drifting pebbles and quenches the core mass growth by pebble accretion \citep{Lambrechts2014b,Liu2019b}. Therefore, no matter what the initial masses of the protoplanets are, their final masses cannot grow beyond the pebble isolation mass. A similar conclusion has already been made in \cite{Liu2019b} where they focus on the planet formation around stars of a higher stellar mass range, from $0.1 \Ms$ to $1\Ms$.

\Fg{mass} clearly shows that the masses of planets correlate with the masses of their (sub)stellar hosts. All forming planets around these UCDs are core-dominated with $M_{\rm p} {\lesssim} 2 \Me$. This is consistent with the finding in \cite{Liu2019b} and \cite{Morales2019}, where gas-dominated planets can only form in systems around stars of ${\gtrsim} 0.2{-}0.3 \Ms$. In the case of $\gamma{=}1$ where protoplanets are born in self-gravitating disks, the planets around $10 \ M_{\rm Jup}$ brown dwarfs can finally grow up to $0.1{-}0.2 \Me$. Assuming that the disk mass is $10 \%$ of the central host, such a planet mass is equivalent to ${\sim} 10\%$ of total solids in disks around these brown objects. The protoplanets around  $0.1 \Ms$ stars maximally reach ${\approx} 2{-}3 \Me$ (Fig. \fgnum{mass}a and \fgnum{mass}c).

In the case of $\gamma {\ne}1$ where protoplanets are born in non-self-gravitating disks, we find in Fig. \fgnum{mass}b and \fgnum{mass}d that in the end planets can reach $1 \Me$  around $0.1\Ms$ stars. However, the protoplanets around $10 \ M_{\rm Jup}$ brown dwarfs fail to grow to Mars-mass. They only turn into planets of $ {\sim} 10^{-3}{-}10^{-2} \Me$. There are two reasons for forming such low-mass planets around brown dwarf disks. First, the initial protoplanet-to-host mass ratio ($q_{\rm p} {=} M_0/M_{\star}$) is lower in brown dwarf disks than that in M dwarf disks. For instance, in the ice line scenario $M_0 {\sim}5 {\times}10^{-7} \Me$ around $10 \ M_{\rm Jup}$ brown dwarfs and  $M_0 {\simeq} 10^{-5} \Me$ around $0.1\Ms$ stars  (magenta solid line in \fg{plt}b). Protoplanets accrete pebbles more slowly when $q_{\rm p}$ is lower \citep{Visser2016}. Second, disks around less massive hosts also evolve faster ($\dot M_{\rm g}$ drops more rapidly, \fg{diskm}). Thus, the solids available for pebble accretion around brown dwarfs also drain out more quickly compared to that for M dwarfs. Altogether, these factors limit the growth of planets above $10^{-2} \Me$ around $10 \  M_{\rm Jup}$ brown dwarfs in non-self-gravitating disks.

For the water-ice line scenario of $r_0{=}r_{\rm ice}$, planets generally contain ${\gtrsim} 15\%$ water. We also find in \fg{mass}a that planets around  low-mass M dwarfs contain less water compared to those around brown dwarfs.  This is because protoplanets around M dwarfs grow faster and are able to migrate inside of the ice line to accrete dry pebbles (\fg{growth}b). Therefore, their final water contents become lower around such stars. As can be seen in Figure 7b of \cite{Liu2019b}, this trend also extends to the systems around solar-mass stars.   However, in non-self-gravitating disks ($\gamma{\ne}1$), the initial masses of protoplanets are at least two orders of magnitude lower than self-gravitating disks (\fg{plt}b). In this situation protoplanets grow slowly, and the inward movement of the ice line is faster than their migration, both for systems around M-dwarfs and brown dwarfs.  Eventually all these planets are substantially water-rich (\fg{mass}b).   For the log-uniformly distributed scenario, since the ice lines are further out around more massive M-dwarfs, a higher fraction of protoplanets around such stars are born inside of the water-ice lines and finally grow into water-deficient planets. This also can be seen in Fig. \fgnum{mass}c and \fgnum{mass}d.

\begin{figure*}[t]
 \includegraphics[scale=0.8, angle=0]{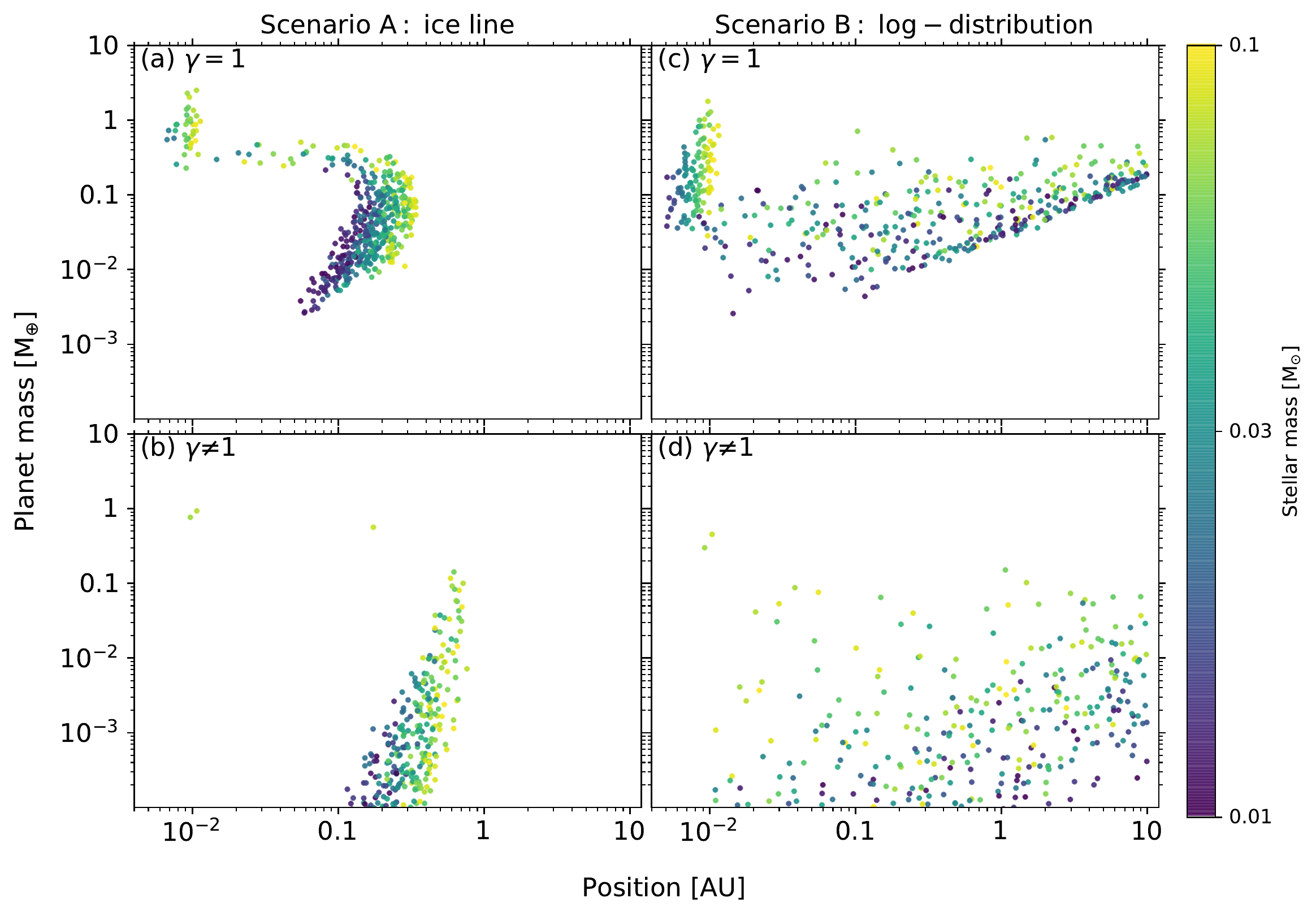}
       \caption{ 
  Monte Carlo sampling plot of the final planet mass vs the semi-major axis, with the ice line planet formation model (Scenario A) in the left, the log-uniform distributed planet formation model (Scenario B) in the right, the self-gravitating disks ($\gamma{=}1$) in the top and the non-self-gravitating disks ($\gamma {\ne} 1$) in the bottom.  The color corresponds to the (sub)stellar mass.  Scenario A only produces close-in planets,  while planets formed in Scenario B have a wide range of orbital distances. Growth and migration are more significant for the protoplanets around low-mass M-dwarfs than those around brown dwarfs.}
\label{fig:rplanet}
\end{figure*}

\fg{rplanet} illustrates the Monte Carlo plot of the planet mass and semi-major axis. The color represents the mass of the ultra-cool dwarf. We can see that short-period planets form in the ice line scenario, whereas planets end up with a wide range of orbital distances in the log-uniformly distributed scenario.
On the other hand, we find that protoplanets around brown dwarfs are likely to grow nearly in-situ, whereas a higher fraction of protoplanets around low-mass M-dwarfs grow substantially and migrate to the inner disk regions. As mentioned before, this is because the disk mass drops rapidly around brown dwarfs, and therefore the growth is marginally limited in this circumstance.

\section{Conclusion}
\label{sec:conclusion}

In this paper we presented  a pebble-driven population synthesis model  to study how planets form around very low-mass stars and brown dwarfs in the (sub)stellar mass range of $0.01 \Ms$ and $0.1 \Ms$. Here we improve our model with respect to \cite{Liu2019b}  by
\begin{itemize}
\item adopting an observed fitting formula of $\dot M_{\rm g}(M_{\star},t)$  from \cite{Manara2012},
\item using  the stellar luminosity $L_{\star}(M_{\star},t)$ calculated from the advanced stellar evolution model of \cite{Baraffe2003,Baraffe2015},
\item choosing realistic initial masses for the protoplanets based on the extrapolation of streaming instability simulations.  
\end{itemize}
The planet formation processes we incorporate are pebble accretion onto planetary cores, gas accretion onto their envelopes, and  type I and type II planet migration (\se{model}). 


 Two hypotheses on the disk conditions during planetesimal formation have been explored, assuming that the disk is either self-gravitating ($\gamma{=}1$), or  non-self-gravitating ($\gamma {\ne}1$). Also, two scenarios for the birth locations of the protoplanets are investigated: either  protoplanets form only at the water ice line or they are log-uniformly distributed over the entire disk. We performed Monte Carlo simulations and the resulting planet population with their final planet masses, water fractions and semi-major axes are presented in \se{pps}.

The key conclusions of this paper are listed as follows: 
\begin{enumerate}[1.]
     \item Disks around low-mass stars and brown dwarfs evolve more rapidly than disks around high mass stars, as  inferred from measured disk accretion rates of \cite{Manara2012}. When disks are younger than $1$ Myr, the disk accretion rate linearly scales with the host's mass. However, as disks evolve over $1$ Myr, $\dot M_{\rm g}$ becomes gradually proportional to $M_{\star}^2$ (\fg{diskm}).
      \item  We derive the characteritic mass of the planetesimals in \eq{Mpl3} based on the extrapolation of the literature streaming instability simulations.  Assuming that the protoplanets form from the planetesimal populations by streaming instability, their masses depend on both radial distances and masses of their hosts (\eq{Mem_new}).   Protoplanets have lower masses when they are born at closer-in orbits and/or around lower-mass hosts, whereas protoplanets are more massive when they form further out and/or around higher mass hosts (\fg{plt}).
        \item Planets formed at short orbital distances ($r {\lesssim} r_{\rm ice}$) can grow their mass substantially.  However, the growth of protoplanets is very limited when they form further out (\fg{growth} and \fg{rplanet}). 
       \item A linear mass correlation between planets and hosts holds when protoplanets form with relative high masses in self-gravitating disks.  However, this correlation breaks down when protoplanets grow in initially non-self-gravitating disks. In this case these protoplanets cannot reach the pebble isolation mass.  Mars-mass planets can form around  $0.01 \Ms$ (${\approx} 10 \  M_{\rm Jup}$) brown dwarfs when massive protoplanets form at early the self-gravitating phase, while protoplanets only grow up to $10^{-3}{-}10^{-2} \Me$ when they form later in the non-self-gravitating phase. For systems around  late M-dwarfs of $0.1 \Ms$, the protoplanets can grow up to $2{-}3\Me$.  All forming planets around these ultra-cool stars are core-dominated with negligible hydrogen and helium gaseous envelopes (\fg{mass}).  
        \item Protoplanets originated at the water-ice line finally grow into water-rich planets, with ${\gtrsim}15\%$ mass in water. Planets formed around higher-mass stars generally end up with lower water fractions. Water-deficient planets ($f_{\rm H_2O}{<}1\%$) can only form when protoplanets are born interior to $r_{\rm ice}$ (\fg{mass}).            
\end{enumerate}

\begin{acknowledgements}
We thank Adam Burgasser, Thomas Ronnet, Urs  Sch\"{a}fer and Rixin Li for useful discussions.
 We also thank the anonymous referee for their useful suggestions and comments. 
B.L. is supported by the European Research Council (ERC Consolidator Grant 724687-PLANETESYS) and the Swedish Walter Gyllenberg Foundation.  M.L. is funded by the Knut and Alice Wallenberg Foundation (Wallenberg Academy Fellow Grant 2017.0287). A.J. thanks the European Research Council (ERC Consolidator Grant 724687-PLANETESYS), the Knut and Alice Wallenberg Foundation (Wallenberg Academy Fellow Grant 2012.0150) and the Swedish Research Council (Project Grant 2018-04867) for research support.
I.P. acknowledges support from an NSF Astronomy \& Astrophysics Research Grant (ID: 1515392). 
T.H. acknowledges support from the European Research Council under the Horizon 2020 Framework Program via the ERC Advanced Grant Origins 832428.

\end{acknowledgements}

\appendix

\section{Disk model equations}
\label{ap:diskmodel}
Here we provide the key equations for our adopted disk model, which includes an inner viscously heated region and an outer stellar irradiation region. The detailed descriptions and derivations are referred to as \cite{Liu2019b}. The gas surface density and disk aspect ratio are given by
  \begin{equation}
 \frac{\Sigma_{\rm g}}{ \rm g \ cm^{-2}}  =
  \begin{cases}
 {\displaystyle   75
    \left( \frac{\dot M_{\rm g}}{10^{-9} \Msyr} \right)^{1/2}  \left(\frac{M_{\star}}{0.1 \ M_{\odot}} \right)^{1/8} } \\
   {\displaystyle  \left(\frac{r}{0.1 \AU} \right)^{-3/8}  }
     \hfill  [\mbox{vis}],  \vspace{0.6cm}\\
 {\displaystyle  250 \left( \frac{\dot M_{\rm g}}{10^{-9} \rm \Msyr} \right)
\left(\frac{M_{\star}}{0.1 \ M_{\odot}} \right)^{9/14}} \\
{\displaystyle  \left(\frac{L_{\star}}{0.01 \ L_{\odot}} \right)^{-2/7}  \left(\frac{r}{0.1 \AU} \right)^{-15/14} }
  \hfill  [\mbox{irr}], 
\end{cases}
\label{eq:sigma}
\end{equation}
and
  \begin{equation}
h = \begin{cases}
 {\displaystyle   0.045
\left( \frac{\dot M_{\rm g}}{10^{-9} \Msyr} \right)^{1/4}
\left(\frac{M_{\star}}{0.1 \ M_{\odot}} \right)^{-5/16}} \\
   {\displaystyle  \left(\frac{r}{0.1 \AU} \right)^{-1/16}  }  
     \hfill  [\mbox{vis}],  \vspace{0.6cm}\\
 {\displaystyle   0.0245
    \left(\frac{M_{\star}}{0.1 \ M_{\odot}} \right)^{-4/7} 
    \left(\frac{L_{\star}}{0.01 \ L_{\odot}} \right)^{1/7}  }\\
     {\displaystyle    \left(\frac{r}{0.1 \AU} \right)^{2/7} } 
       \hfill  [\mbox{irr}],
\end{cases}
\label{eq:aspect}
\end{equation}
where the top and bottom rows represent the quantities in the inner viscously heated region and outer stellar irradiation region, respectively,  $\dot M_{\rm g}$ is the disk accretion rate, $M_{\star}$, $L_{\star}$ are (sub)stellar mass and luminosity,  and $r$ is the radial distance to the central star.

The disk transition radius between these two regions is given by
\begin{equation}
      \begin{split}
   r_{\rm tran} = & 0.56
    \left( \frac{\dot M_{\rm g}}{10^{-9} \rm \Msyr} \right)^{28/39}
    \left(\frac{M_{\star}}{0.1 \ M_{\odot}} \right)^{29/39} \\
& \left(\frac{L_{\star}}{0.01 \ L_{\odot}} \right)^{-16/39}    \AU.
     \end{split}
     \label{eq:rtrans}
\end{equation}
The location of the water-ice line is calculated by equating the saturated pressure and $\rm H_2 O$ vapor pressure (Eq. 35 of \cite{Liu2019b}). The water-ice line in different disk regions can be approximately given by   
  \begin{equation}
\frac{r_{\rm ice}}{\AU} = \begin{cases}
 {\displaystyle  0.26
    \left( \frac{\dot M_{\rm g}}{10^{-9} \rm \  M_{\odot}yr^{-1}} \right)^{4/9}
    \left(\frac{M_{\star}}{0.1 \ M_{\odot}} \right)^{1/3}   }  
     \hfill  [\mbox{vis}],\\
 {\displaystyle  0.075 \left(\frac{M_{\star}}{0.1 \ M_{\odot}} \right)^{-1/3} 
    \left(\frac{L_{\star}}{0.01 \ L_{\odot}} \right)^{2/3}}
        \hfill  [\mbox{irr}].
\end{cases}
\label{eq:rice}
\end{equation}
The ice line location is the maximum of these two values (  $r_{\rm ice} {= }\max \left[r_{\rm ice,vis},r_{\rm ice,irr} \right]$).


 Inserting \eq{rice} into \eq{aspect}, the disk aspect ratio at the ice line location is 
  \begin{equation}
\begin{split}
h_{\rm ice} = \begin{cases}
{\displaystyle 0.049 \left( \frac{\dot M_{\rm g}}{10^{-9} \Msyr} \right)^{2/9} \left(\frac{M_{\star}}{0.1 \ M_{\odot}} \right)^{-1/3}} 
 \hfill  [\mbox{vis}],\\
 {\displaystyle 0.023 \left(\frac{M_{\star}}{0.1 \ M_{\odot}} \right)^{-2/3}  \left(\frac{L_{\star}}{0.01 \ L_{\odot}} \right)^{1/3}}
 \hfill  [\mbox{irr]}.
\end{cases}
    \end{split}
    \label{eq:hice}
\end{equation}

The gas surface density at the ice line and $10$ AU  are also derived, 
\begin{equation}
\frac{\Sigma_{\rm g}}{ \rm g \ cm^{-2}} = \begin{cases}
 {\displaystyle  52
    \left( \frac{\dot M_{\rm g}}{10^{-9} \Msyr} \right)^{1/3}  }  
 \hfill    [\mbox{r = } r_{\rm ice}] ,     \vspace{0.6cm} \\ 
 {\displaystyle  1.8 \left( \frac{\dot M_{\rm g}}{10^{-9} \rm \Msyr} \right)
\left(\frac{M_{\star}}{0.1 \ M_{\odot}} \right)^{9/14}} \\
{\displaystyle  \left(\frac{L_{\star}}{0.01 \ L_{\odot}} \right)^{-2/7} }
\hfill  [\mbox{r = 10   \rm AU}].
\end{cases}
\label{eq:sigma_2}
\end{equation}

\section{Evolution of stellar luminosity}
\label{ap:luminosity}
\begin{figure}[t]
	     \includegraphics[width=\columnwidth]{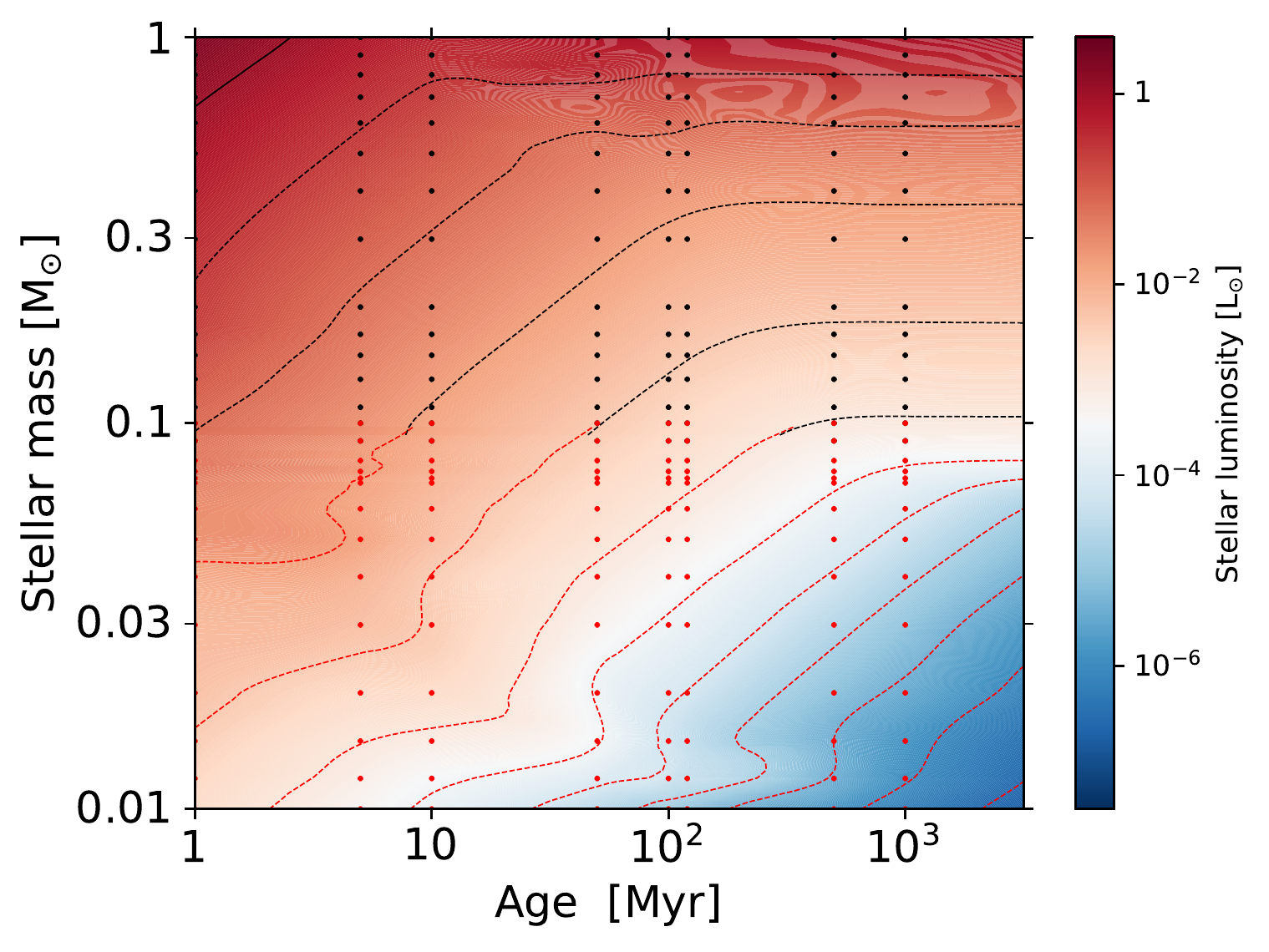}
		\caption{  Stellar luminosity as function of time and stellar mass based on the stellar evolutionary model of \cite{Baraffe2003,Baraffe2015}.  The dots are data from the stellar evolution model, and the dashed line corresponds to the same luminosity. The black and red represent the stars above or below the  hydrogen-burning limit.  When the star is younger than $\sim 10$ Myr, $L_{\star} \propto M_{\star}^{1-2}$, while $L_{\star}  \propto M_{\star}^{3-4}$ when stars are older.
		}
	\label{fig:Ls}
	\end{figure}

Stars/Brown dwarfs of various masses evolve differently, and their luminosities vary with the stellar type and evolutional stage. We present an interpolation calculation of $L_{\star} (M_{\star}, t)$ based on the stellar evolution model of \cite{Baraffe2003} (below the hydrogen-burning limit) and \cite{Baraffe2015} (above the hydrogen-burning limit).  It is worth noting that \cite{Baraffe2015} also contains the evolutional tracks of stars and brown dwarfs with the mass range of $0.01\Ms$ and $0.1 \Ms$. However, in their table the evolution of low-mass brown dwarfs stop early, \eg,  $0.01  \Ms$ brown dwarfs at $10$ Myr.   \cite{Baraffe2003} contains a much long-term evolution for such low-mass objects. We compare the overlapping parameter space and found that the deviation of $L_{\star}$ at certain $M_{\star}$ and $t$  from the above two studies is within $35\%$, relatively small compared to a few orders of magnitude decreasing of $L_{\star}$ over time. In a future follow-up study, we plan to investigate the long-term evolution and atmosphere water-loss of these terrestrial planets.  We therefore combine these two datasets together. 

\Fg{Ls} shows a full map of stellar luminosities as functions of stellar masses and time. The (sub)stellar mass ranges from $10 \ M_{\rm Jup}$  to solar-mass.  The dots represent the results from stellar evolution models, where the black and red correspond to stars in the mass range above or below the hydrogen-burning limit, \ie , $M_{\star} = 0.08 \Ms$. As pointed out in \se{stellar}, the  stars less than $\sim 10$ Myr show a  $L_{\star}-M_{\star}^{1-2}$ relation. After a few tens of Myr, their luminosities behave as   $L_{\star} \propto M_{\star}^{3-4}$.

\section{Model assumption exploration}

In order to test how the results in \se{pps} vary with different model assumptions, we conduct two additional sets of simulations.  Different treatments of $\dot M_{\rm g}{-}M_{\star}$ relation and the particle size are explored in \aps{mdot}{size}, respectively. 

\subsection{Different disk accretion rate-host mass correlation}
\label{ap:mdot}

In the main paper we adopt $\dot M_{\rm g} (M_{\star}, t)$ from \cite{Manara2012}. Here we choose different $\dot M_{\rm g}$ prescription from literature work. For instance, \cite{Manara2017} obtained a correlation of $\dot M_{\rm g}{-}M_{\star}^{2.3}$ (their Eq.(5)) for stars around  Chamaeleon I star formation region. But in their work there are no inferred ages of  young (sub)stellar objects. Although the mean age of Chamaeleon I cluster is ${\sim}2$ Myr, the age spreading can be as large as $2$ Myr among individual objects. With no better information, we assume $\dot M_{\rm g} {\propto}t^{-3/2}$ from the self-similar solution of the viscous accretion disk. But again, by this setup the evolution of disks around various mass stars/brown dwarfs are the same, which is the key difference from \cite{Manara2012}. Nevertheless, from this exploration, we can learn the impact of different disk accretion and evolution conditions on forming planets. We also assume a constant $\dot M_{\rm g}$ in the early embedded phase of $t{\leq} 0.3$ Myr, to avoid an unrealistically high accretion when $t $ approaches $0$ yr.

\begin{figure}[t]
	     \includegraphics[width=\columnwidth]{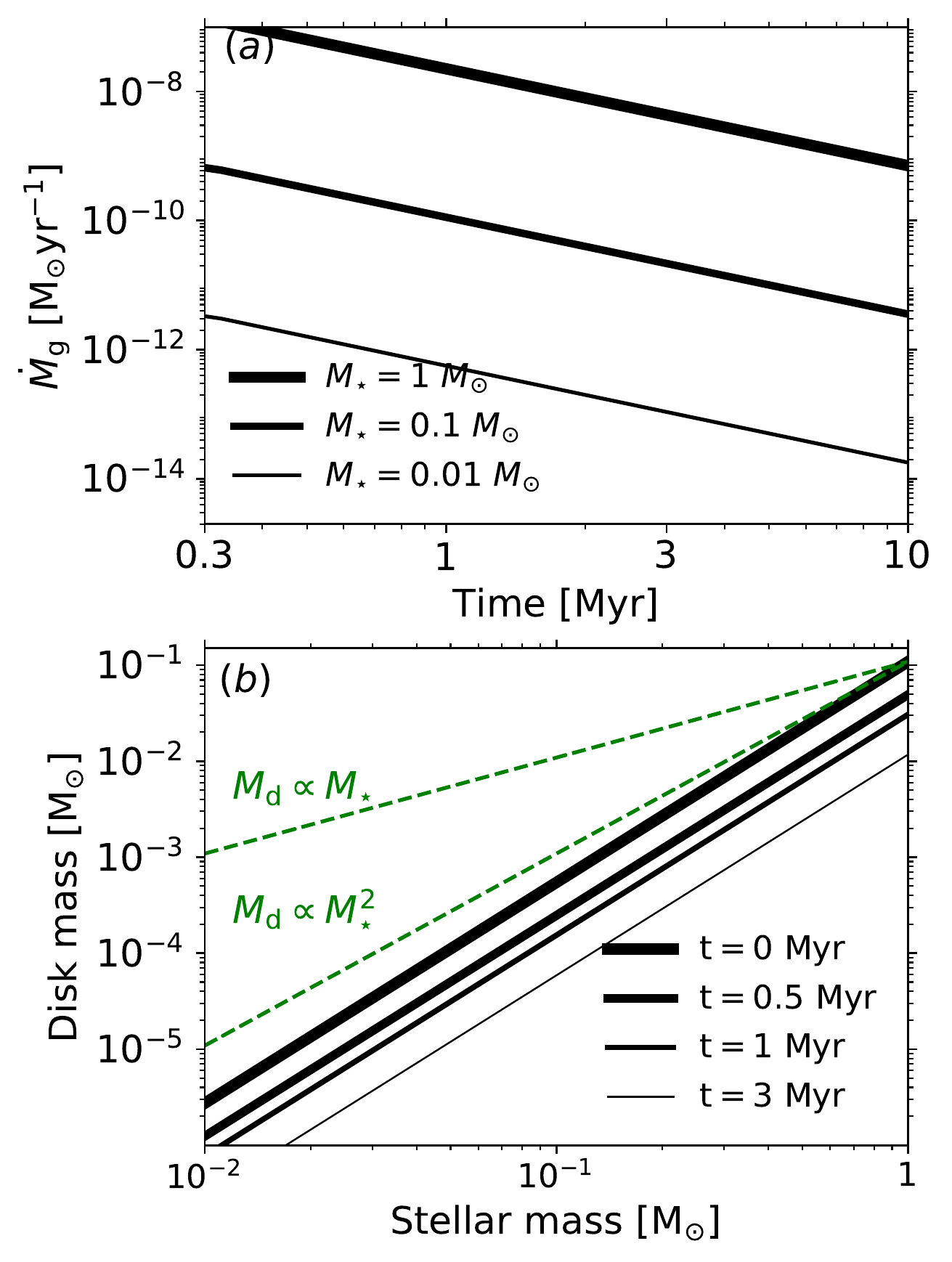}
		\caption{  Panel(a): time evolution of disk accretion rates among three different mass hosts, based on Eq. (5) of \cite{Manara2017}. 
		Panel (b): time and host's mass dependences on disk mass based on Eq.(5) of \cite{Manara2017}. The solid lines from thick to thin represent the ages of systems, from $0$, $0.5$, $1$ to $3$ Myr.  The green dashed lines correspond to the linear, quadratic relation between the disk mass and the central host's mass. 
		The time evolution of the disks is the same around all mass stars.  }
	\label{fig:diskm17}
	\end{figure}
	
	\begin{figure*}[tbh]
 \includegraphics[scale=0.8, angle=0]{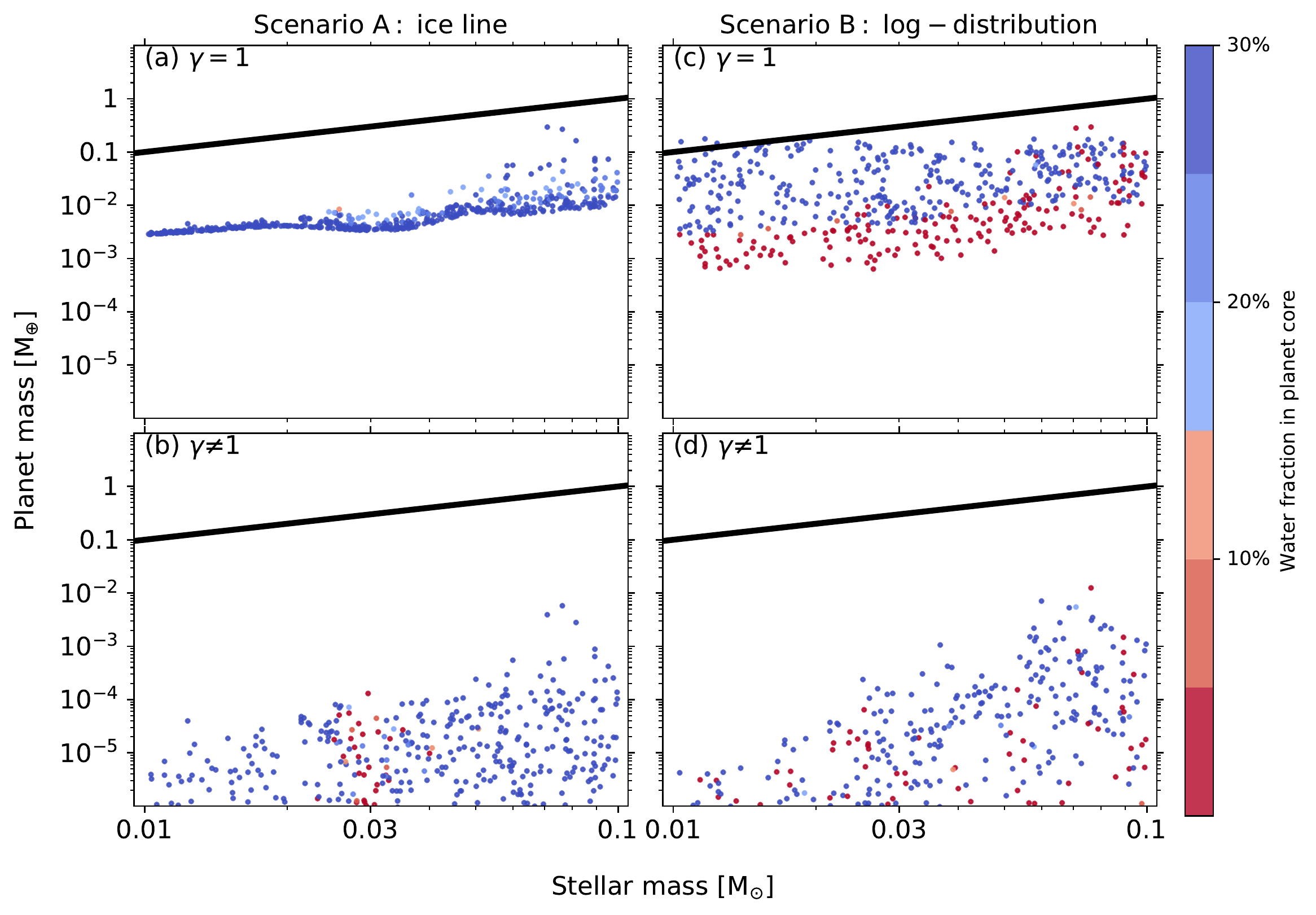}
       \caption{  
 Monte Carlo sampling plot of the planet mass vs the stellar mass, with the ice line planet formation model (Scenario A) in the left, the log-uniform distributed planet formation model (Scenario B) in the right, the self-gravitating disks ($\gamma{=}1$) in the top and the non-self-gravitating disks ($\gamma {\ne} 1$) in the bottom. The initial disk conditions are followed by Eq. 5 of \cite{Manara2017}. The color corresponds to the  water mass fraction in the planetary core.  The black line represents a constant planet-to-star mass ratio of $3 {\times}10^{-5}$ motivated by \cite{Pascucci2018}. 
The planet growth is significantly suppressed compared to \fg{mass}.  Protoplanets cannot grow their masses in disks around brow dwarf disks of $<0.05 \Ms$ due to very low disk masses.
    }
\label{fig:massdisk}
\end{figure*} 
	
	Compared to \fg{diskm}, the time and host's mass dependences on $\dot M_{\rm g}$ and $M_{\rm d}$ are also exhibited in \fg{diskm17} based on Eq. (5) of  \cite{Manara2017}.  We find that in this case disk masses around low-mass stars and brown dwarfs are much lower than those in \fg{diskm}. In particular, disks around  $0.01 \Ms$ brown dwarfs are less than $0.1 \%$ of the masses of their central hosts.

	\Fg{massdisk} shows the resulting planets from such disk conditions.  Comparing \fg{massdisk} and \fg{mass}, we find that the final planet populations differ significantly when the disk condition varies.  Only the protoplanets close to the ice line locations in disks around hosts of ${\gtrsim}0.05\Ms$ can grow their masses and migrate to the inner disk regions. 
	However, protoplanets seldom grow in disks around hosts of ${\lesssim}0.05\Ms$ due to their low disk accretion rates.  As expected, the building block materials provided for planet formation is much less in such disks. The final planet masses and semi-major axes remain quite close to their initial values.  For systems around $0.01 \Ms$ brown dwarfs, planets of Mars-mass can form only at further out disk locations in self-gravitating disks. Different from \fg{mass},  there is no a clear linear correlation between $ M_{\rm p}$ and $M_{\star}$ in \fg{massdisk}.

 	 \cite{Manara2017} also provided two power-law fitting formula for the disk accretion rate (their Eq. 4). If this can indeed be extended to the brown dwarf mass regime, surprisingly, it indicates that very low mass brown dwarfs have no accreting disks (extremely low $\dot M_{\rm g}$).  After forming protoplanets by streaming instability, the subsequent planet growth cannot proceed in those circumstances.

 To summarize, how protoplanetary disks accrete and evolve with time is a crucial condition to planet formation. Planet growth is inhibited in brown dwarf disks when accretion rates are lower than ${<}10^{-13} \Msyr$.

\subsection{Particles limited by fragmentation, bouncing and radial drift}
\label{ap:size}
In the main paper, we set the particles to be all one-millimeter in size, which is mainly motivated by disk observations. In this subsection, we consider the sizes of particles from a theoretical point of view. 
Apart from the aforementioned bouncing barrier, dust growth also faces the radial drift and fragmentation barriers.  In \cite{Birnstiel2012}'s dust coagulation model,  grain growth is limited by radial drift in the outer disk regions, whereas in inner disk regions fragmentation dominates. They also analytically derived the maximum stokes numbers in these two regimes.  

The stokes number of particles in the fragmentation limit is $\tausf {=} v_{\rm frag}^2/\alphat c_{\rm s}^2$, where  $v_{\rm frag}$ is the fragmentation threshold velocity, approximately $3 \  \rm  m/s$ for silicate aggregates \citep{Blum1993,Blum2008} and $ 10 \  \rm  m/s$ for ice aggregate \citep{Gundlach2015}. The above equation assumes that particles' relative velocities are lead by disk turbulence.  The  stokes number in the drift regime is $\tausd {=} \epsilon \Sigma_{\rm peb}/\Sigma_{\rm g} \eta$, where $\epsilon$ represents the order of unity coagulation coefficient. The surface density ratio between pebbles and gas can be calculated based on the constant flux ratio $\xi$ assumption, which is given by 
\begin{equation}
\begin{split}
\frac{\Sigma_{\rm peb}}{\Sigma_{\rm g}} = & \frac{\dot M_{\rm peb}}{\dot M_{\rm g}} \frac{u_{\rm g}}{u_{\rm g} + v_{\rm d} } = \xi \left[ 1 +  \frac{2 \tausd \eta}{3 \alphag (1 + \tausd^2) h^2}   \right],
   \end{split}
\end{equation}
where $u_{\rm g}$ and $v_{\rm d}$ are the radial drift velocities for gas and dust, respectively. 
 We assume in the bouncing limit that the particles' size is $1 $ mm. 

\begin{figure}[t]
	     \includegraphics[width=\columnwidth]{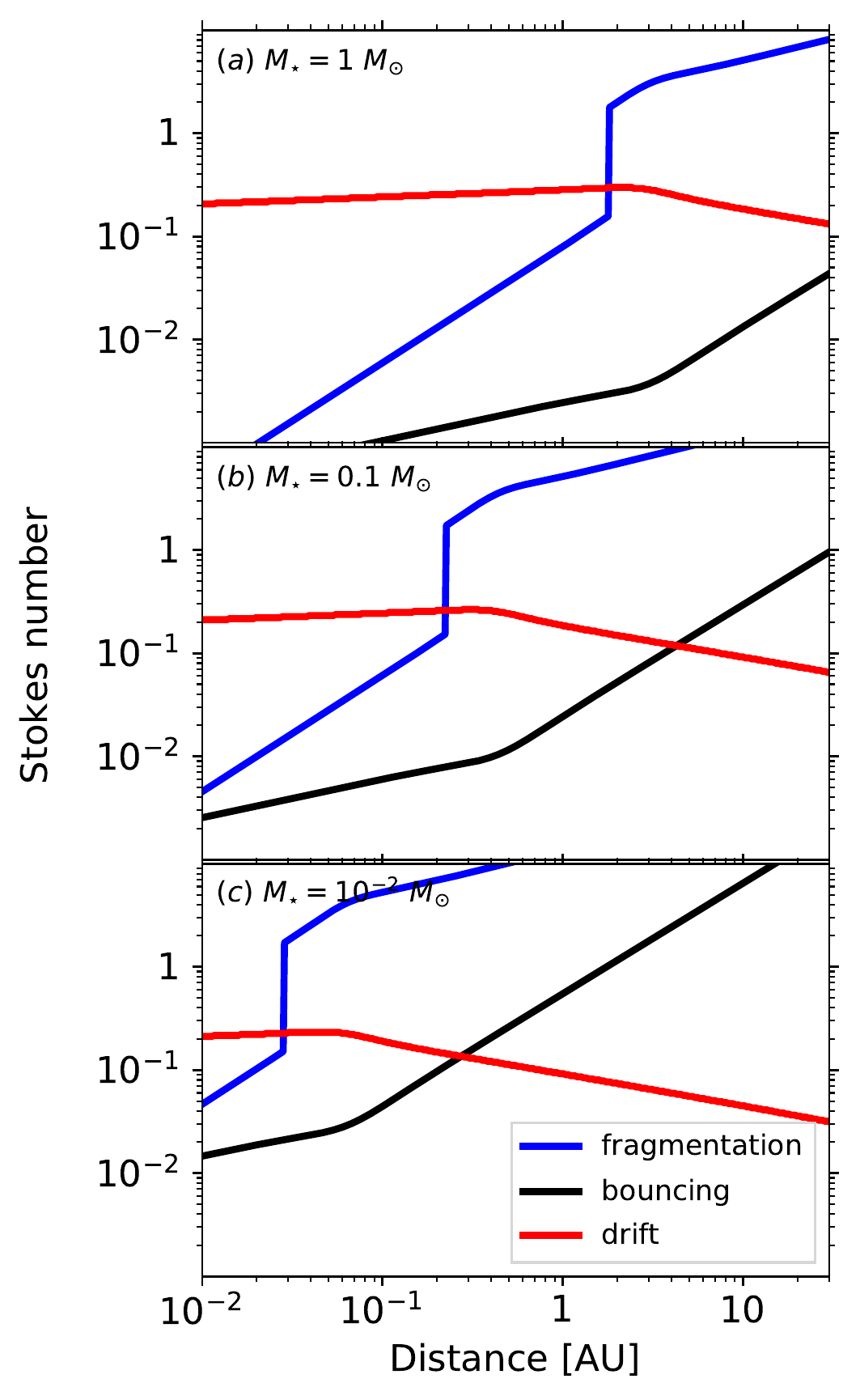}
		\caption{  Stokes number of particles in the fragmentation (blue), bouncing (black) and radial drift (red) limit in $1$ Myr old disks around hosts of solar-mass (upper), $0.1\Ms$ (middle) and $0.01 \Ms$ (lower). The size of the particles is limited by bouncing in disks around solar-mass stars. As masses of central hosts decrease, the radial drift becomes more significant in the outer disk regions. }
	\label{fig:StFBD}
	\end{figure}

\fg{StFBD} shows the stokes numbers of particles in the radial drift (red), bouncing (black) and fragmentation (blue) limits in disks around solar-mass stars (upper), $0.1\Ms$ stars (middle) and $0.01 \Ms$ brown dwarfs (lower) at $t= 1$ Myr, respectively. We find that the Stokes number in the bouncing limit is always lower than that in the fragmentation limit in disks around the explored central hosts. For disks around solar-mass stars, $\tausb$ is always lower than $\tausd$. However,  $\tausb{>}\tausd$ in UCD disks when the particles are beyond a few AUs.  This means that millimeter-sized particles drift too fast in these disks. As pointed out by \cite{Pinilla2013}, the retention of such size particles at a few tens of AUs in brown dwarf disks is problematic. 

We calculate the Stokes number to be the minimum value of the above three limits. \Fg{massFBD} exhibits  the population synthesis simulations with the updated particle sizes.  When comparing \fg{massFBD} and  \fg{mass}, we find that the planet population resulting from the this model (varied particle sizes)  is very similar to the model (fixed particle sizes) presented in \se{pps}.  This is because, firstly, in the inner disk region of $r{\lesssim} r_{\rm ice}$, the size of pebbles is anyway set by the bouncing barrier, making these two models no difference. Secondly, in the outer disk region, pebbles are in the radial drift regime with a very low Stokes number.  In this case, the pebble accretion is in the inefficient $3$D regime, and therefore the protoplanets are hard to grow their masses above  $10^{-2} \Me$ (panel (d) in \fg{massFBD}). However, for the one-millimeter-sized particle model, their Stokes number soon becomes much larger than unity (\fg{St}) and pebble accretion is quickly quenched. Therefore, in that case, protoplanets are also difficult to grow large.

\begin{figure*}[tbh]
 \includegraphics[scale=0.8, angle=0]{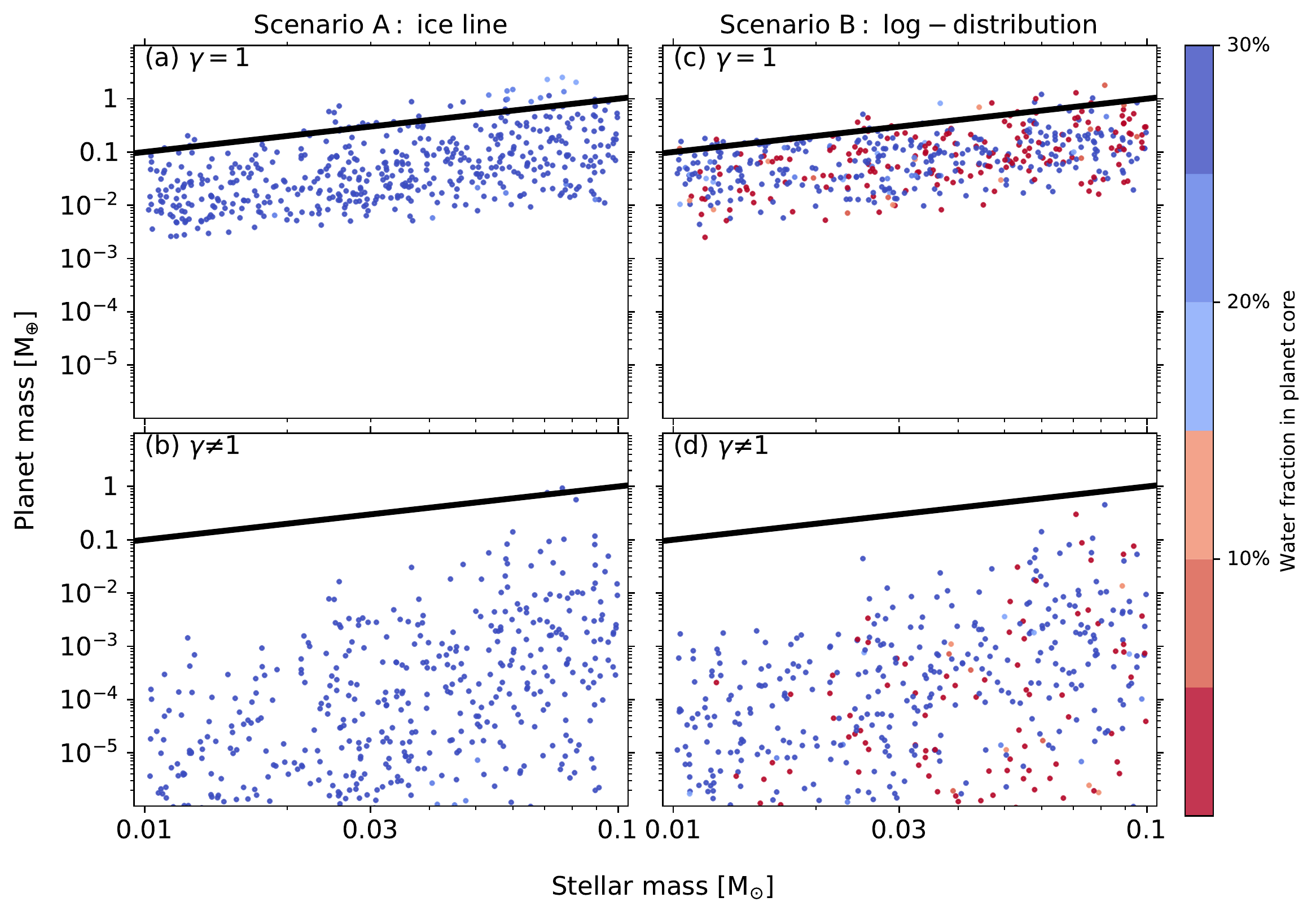}
       \caption{  
 Monte Carlo sampling plot of the planet mass vs the stellar mass, with the ice line planet formation model (Scenario A) in the left, the log-uniform distributed planet formation model (Scenario B) in the right, the self-gravitating disks ($\gamma{=}1$) in the top and the non-self-gravitating disks ($\gamma {\ne} 1$) in the bottom.  The color corresponds to the  water mass fraction in the planetary core.  The black line represents a constant planet-to-star mass ratio of $3 {\times}10^{-5}$ motivated by \cite{Pascucci2018}. Here we set the particle sizes by combined effects of radial drift, bouncing, and fragmentation. The resulting planet population is very similar to that of the fixed-sized particles model in \fg{mass}. 
    }
\label{fig:massFBD}
\end{figure*}

\bibliographystyle{aa}
\bibliography{reference}

\end{document}